\documentclass[%
aip,
pop,
reprint%
]{revtex4-1}

\usepackage{hyperref}
\hypersetup{
    colorlinks=true,
    citecolor=blue,
    linkcolor=blue,
    filecolor=blue,      
    urlcolor=blue,
}

\usepackage{color} 

\usepackage{graphicx}

\usepackage{amsmath}
\usepackage{amsfonts}
\usepackage{amssymb}
\usepackage{amsthm}

\usepackage{siunitx}

\newcommand{\bfA}{{\mathbf{A}}}

\newcommand{\bfE}{{\mathbf{E}}}
\newcommand{\bfe}{{\mathbf{e}}}

\newcommand{\bfH}{{\mathbf{H}}}
\newcommand{\bfJ}{{\mathbf{J}}}

\newcommand{\bfr}{{\mathbf{r}}}

\newcommand{\rmc}{{\mathrm c}}
\newcommand{\rmd}{{\mathrm d}}
\newcommand{\upd}{{\, \mathrm d}}
\newcommand{\rme}{{\mathrm e}}
\newcommand{\upe}{{\, \mathrm e}}
\newcommand{\rmg}{{\mathrm g}}
\newcommand{\rmi}{{\mathrm i}}

\newcommand{\rmr}{{\mathrm r}}
\newcommand{\bcA}{{\boldsymbol{\mathcal A}}}
\newcommand{\cA}{{\mathcal A}}

\newcommand{\cE}{{\mathcal E}}
\newcommand{\bcE}{{\boldsymbol{\mathcal E}}}

\newcommand{\cG}{{\mathcal G}}
\newcommand{\cH}{{\mathcal H}}
\newcommand{\bcH}{{\boldsymbol{\mathcal H}}}
\newcommand{\cI}{{\mathcal I}}
\newcommand{\cK}{{\mathcal K}}

\newcommand{\cM}{{\mathcal M}}

\newcommand{\cR}{{\mathcal R}}
\newcommand{\cS}{{\mathcal S}}
\newcommand{\cV}{{\mathcal V}}

\newcommand{\RR}{{\mathbb R}}

\newcommand{\VV}{{\mathbb V}}
\newcommand{\ZZ}{{\mathbb Z}}

\newcommand{\sK}{{\mathsf K}}

\newcommand{\frK}{{\mathfrak{K}}}
\newcommand{\frk}{{\mathfrak{k}}}
\newcommand{\frH}{{\mathfrak{H}}}

\newcommand{\derivep}[2]{ \frac{\partial #1}{\partial #2} }

\newcommand{\sinc}{{\mathrm{sinc}}}

\usepackage{fancyhdr}
\begin{document}

\title{
Time simulation of the nonlinear wave-particle interaction in meters long traveling-wave tubes 
}

\author{Damien~F.~G.~Minenna}%
\affiliation{CEA, DAM, DIF, F-91297 Arpajon, France}
\affiliation{Centre National d'{\'E}tudes Spatiales, F-31401 Toulouse, France}%
\affiliation{Thales AVS/MIS, 78140 V\'elizy, France}
\affiliation{Aix-Marseille Universit{\'e}, CNRS, PIIM UMR 7345, F-13397 Marseille, France}
\author{Khalil~Aliane}%
\affiliation{Centre National d'{\'E}tudes Spatiales, F-31401 Toulouse, France}%
\affiliation{Thales AVS/MIS, 78140 V\'elizy, France}
\affiliation{Aix-Marseille Universit{\'e}, CNRS, PIIM UMR 7345, F-13397 Marseille, France}
\author{Yves~Elskens}%
\email[Corresponding author: ]{yves.elskens@univ-amu.fr}
\affiliation{Aix-Marseille Universit{\'e}, CNRS, PIIM UMR 7345, F-13397 Marseille, France}
\author{Alexandre~Poy{\'e}}%
\affiliation{Aix-Marseille Universit{\'e}, CNRS, PIIM UMR 7345, F-13397 Marseille, France}
\author{Fr{\'e}d{\'e}ric~Andr{\'e}}%
\affiliation{Thales AVS/MIS, 78140 V\'elizy, France}
\author{J{\'e}r\^{o}me~Puech}%
\affiliation{Centre National d'{\'E}tudes Spatiales, F-31401 Toulouse, France}
\author{Fabrice~Doveil}%
\affiliation{Aix-Marseille Universit{\'e}, CNRS, PIIM UMR 7345, F-13397 Marseille, France}

\date{Submitted: 8 June 2021 . Accepted: 17 August 2021. Published Online: 14 September 2021}


\begin{abstract} 
We propose a multi-particle self-consistent Hamiltonian (derived from an $N$-body description) that is applicable for periodic structures such as traveling-wave tubes (TWTs), gyrotrons, free-electron lasers, or particle accelerators. We build a 1D symplectic multi-particle algorithm to simulate the nonlinear wave-particle interaction in the time domain occurring in an experimental 3-meters long helix TWT. Our algorithm is efficient thanks to a drastic reduction model. A 3D helix version of our reduction model is provided. Finally, we establish an explicit expression of the electromagnetic power in the time domain and in non-monochromatic (non ``continuous waveform'') regime.
\end{abstract}

\maketitle
\thispagestyle{fancy}

\section{Introduction}

Large-scale simulations in electrodynamics or plasma physics are challenging, especially when investigating nonlinear multi-scale effects. One possible approach, the $N$-body (many-body) description\cite{els03,esc18}, consists in taking into account the dynamics of all charged particles. Sadly, this approach is impractical due to its excessive number of degrees of freedom (DOFs) of fields and particles, especially to model full centimeters to meters long devices with billions of particles.

However, for periodic structures, including traveling-wave tubes (TWTs), backward-wave tubes, gyrotrons, gyro-TWTs, extended interaction klystrons, free-electron lasers or particle accelerators, one can tame the number of DOFs
by using a reduction model called the Kuznetsov discrete model.\cite{kuz80,rys07,rys07ieee,rys09,ber11,ber11b,the16,ter17,min19ps,rys21} 
The discrete model is then combined with an $N$-body self-consistent Hamiltonian description\cite{and13,min17,min18,min19phd} in one space dimension (1D) to model the momentum exchange between fields and particles. From this model reduction, we build a symplectic multi-particle integrator\cite{min19ted} called \textsc{dimoha}, which is designed to preserve the Hamiltonian structure. Its drastic DOF reduction enables fast simulations in the time domain. Thanks to the discrete model, our integrator uses macro-particles, therefore it is more accurately classified as a multi-particle Hamiltonian (or multi-symplectic) algorithm, as others have proposed.\cite{xia13,evs13,sha14,qin16,web16,qia17,qia18} In addition, $N$-body simulations using the real number of particles are feasible for certain low-charge beams.\cite{max13,qia07}
The $N$-body approach is also used in textbooks 
to easily introduce elementary electrodynamics behaviours,\cite{jac99} 
the physics of free-electron lasers\cite{fre92} 
and nonlinear microscopic plasma phenomena\cite{els03,esc18} such as trapping, Landau damping or Hamiltonian chaos.

\begin{figure}
\centering
\includegraphics[width=\columnwidth]{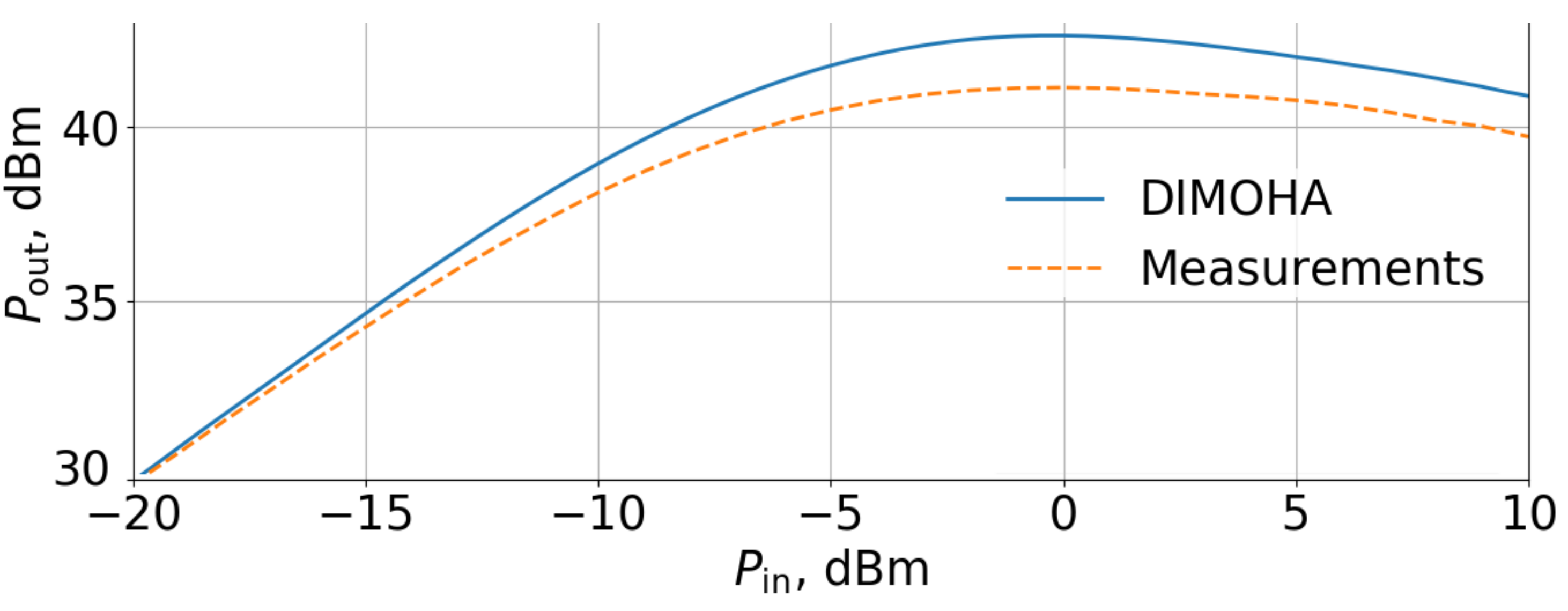}
\caption{
Output power (time average over one period) versus the input power at 8.4 GHz from our algorithm \textsc{dimoha} (Eq.~\eqref{e:powerHarmo}) and measurements.\cite{thales} Those are from one of the two identical (for redundancy) helix TWTs built by Thales that equip the probe New Horizons (NASA).
The power is expressed in dBm from watts as $P [{\rm dBm}] = 10 \log_{10} (P [{\rm W}] \cdot 10^{3})$.
} \label{f:newhor}
\end{figure}

Our approach is applied to simulate the slow-wave structure (SWS) of TWTs,\cite{min19epjh,pie50,gil11} a class of vacuum electron tubes used for basic plasma physics experimental investigation. Plasmas are noisy and can be hard to
control or analyse.
However, the helix slow-wave structure of a TWT can behave analogously to a plasma, allowing to study the self-consistent and nonlinear effects of electrons in plasma waves. 
For this purpose, a huge 3 meter long TWT was built\cite{dim77,dim78,dim82} at the University of California, San Diego in 1976, 
and 4 meter one at Aix-Marseille.\cite{dov06,dov08,dov10,des20}
TWTs are also widely used for space telecommunication.\cite{min19epjh} 
For example, two helix TWTs equip NASA's probe \textit{New Horizons} (flyby of Pluto and Kuiper belt objects) 
to amplify signals before sending them to Earth. 
The two TWTs are powerful enough so that we can receive a signal from more than 50 AU in 2021.
For illustration, Fig.~\ref{f:newhor} displays the comparison of the TWT that equips \textit{New Horizons} against our algorithm \textsc{dimoha}, that is currently adapted for
industrial conception.

This paper has several purposes: 
\begin{itemize}
\item Our main goal is to perform fast and accurate simulations of \textbf{meters long TWTs}. 
This was not yet achieved with a space-time description.
\item Our multi-particle model was introduced briefly in Ref.~\onlinecite{min18},  
which focused on the momentum exchange and monochromatic electromagnetic power.
Since then, we extended the discrete model to allow practical utilisations.
The validation of \textsc{dimoha} against a 15 cm long industrial TWT was presented in Ref.~\onlinecite{min19ted} but without describing our model. 
The second goal of this paper is to \textbf{describe our approach in details and especially how we reduce the number of DOFs} of the system.
To fully understand our approach, 
we provide a complete and pedagogical description about the use of the discrete model in the time domain, 
revisiting past publications.
\item The heart of the discrete model is the field decomposition
separating their time- and space-dependence (see Eqs~\eqref{e:VsnEsn}-\eqref{e:IsnHsn} below). 
\textbf{The time dependences can be computed from a wave-particle interaction model} such as our multi-particle model or with a fluid model.\cite{min19ps}
In addition, we \textbf{extend the discrete model for a 3D helix geometry to compute the space-dependences}.
\item Finally, we investigate the \textbf{electromagnetic power using the discrete model}. We undertake to compute it in time and position using the helix geometry and we propose an explicit expression in non-monochromatic regime.
\end{itemize}

This paper is organized as follows. 
Section~\ref{s:dimodel} recalls basics of the discrete model.  
In Section~\ref{s:1DNbodyHam}, we present our multi-particle self-consistent Hamiltonian (\textsc{dimoha}) model.
In Section~\ref{s:validation}, a validation of our model against measurement from a 3-meters long TWT is presented.
In Section~\ref{s:sheathDIMO}, we present the application of the discrete model for the helix geometry.
Section~\ref{s:power} presents the new power expression with this geometry.
Our main results are summarized in the conclusion. To help readers, brief syntheses are provided at the beginning of each section.

\section{The discrete model}
\label{s:dimodel}

\subsection{Section synthesis}

The discrete model presented is this section is a drastic reduction model for classical electromagnetic field.
Its most important equations are the reformulation of fields Eqs~\eqref{e:VsnEsn} and \eqref{e:IsnHsn} that is exact for periodic structures.
Considering a single mode ($s=0$), fields are now expressed in terms of the product of temporal amplitudes $\cV_{n}$ and $\cI_{n}$ times shape functions $\bcE_{n}$ and $\bcH_{n}$.
One of the goals of this paper is to compute those factors.
The temporal amplitudes can be computed with a wave-particle theory, such as in the next section.
Shape functions depend only on the structures. 1D electric shape functions are provided in the next section, from the dispersion relation and the impedance, while 3D shape functions for the helix geometry are expressed in Section~\ref{s:sheathDIMO}.

In addition  with the reformulation of fields Eqs~\eqref{e:VsnEsn}-\eqref{e:IsnHsn}, that we call the representation per cell, the discrete model also introduces a wavenumber ($\beta$) based representation of fields. Both representations are linked through the Gel'fand transform.

Within the discrete model, the Maxwell equations that describe the field dynamics can be rewritten. The system behaves as an oscillator chain with coupling coefficients from the dispersion relation.

Until Section~\ref{s:sheathDIMO}, the discrete model is presented without imposing a particular geometry (informations on the geometry is encoded in the dispersion relation and in the interaction impedance  that are inputs of our model). 

This section summarises works 
from Refs~\onlinecite{kuz80,rys07,rys07ieee,rys09,
ber11,ber11b,and13,the16,ter17,min17,min18,min19ps,min19phd,rys21} for a time-domain version of the discrete model.
A linear frequency-domain version of the discrete model is described in Ref.~\onlinecite{min19ps}.
Note that the discrete model was also used in Ref.~\onlinecite{min20epjd} to resolve the Abraham-Minkowski dilemma in waveguides: 
there coexist two different field momenta, one expression being canonical and the other one being kinematic.
We also remark that pseudospectral models\cite{con04,set17,set18} share several similarities with the discrete model. 
However, these models work with Fourier transforms, instead of the Gel'fand transform, 
and are based on the telegrapher's equations, instead of Maxwell equations.
A comparison between both models is currently under investigation.\cite{ali21}

\subsection{Gel'fand transform and Helmholtz equations}
\label{s:Gelfand}

The interaction zone (the slow-wave structure (SWS) for TWTs) is considered periodic along the $z$-axis with $d$ the pitch of periods (cells) with cell volume $\VV_0$, 
filling the total volume $\VV_\ZZ = \cup_{n \in \ZZ} \{ \bfr + n d \bfe_z : \bfr \in  \VV_0\}$.
The central feature of the model is the discrete translation invariance along the length of SWS, enabling the Floquet (or Bloch) theorem,
whereby any function $\cG(z)$ can be expanded 
on a basis of functions $\cG_\beta$ satisfying the Floquet condition
\begin{equation}
  \cG_\beta (z + n d) 
  = \cG_\beta (z) \, \rme^{- \rmi n \beta d} \, ,
  \label{e:Floquet}
\end{equation}
with $\beta \in [ - \pi / d, \pi/d]$ the fundamental wavenumber (a.k.a.\ propagation constant) associated with Bloch's theorem 
and $n \in \mathbb{Z}$ labelling the $n$\textsuperscript{th} period (or cell) of the periodic structure.
The Fourier representation of those functions $\cG_\beta$, with $ z$ or $\bfr$ as a parameter, is
\begin{equation}
  \cG_{n} (z) 
  = \frac{1}{2 \pi} \int^{\pi}_{-\pi} \cG_{\beta} (z) \, \rme^{-\rmi n \beta d} \, \rmd (\beta d) \, , 
  \label{e:betainverse}
\end{equation}
where $\cG_{n}$ (in the representation per cell) is given by the Gel'fand $\beta$-transform \cite{kuz80,rys07,rys07ieee} (a.k.a.\ Gel'fand mapping, $kq$-representation or Zak transform) as
\begin{equation}
  \cG_{\beta} (z) =
  \sum_{n \in \ZZ} \cG_{n} (z) \, \rme^{\rmi n \beta d} \, . 
  \label{e:fourierBeta}
\end{equation}
Equation~\eqref{e:betainverse} is sometimes referred to as the Gel'fand inverse $\beta$-transform.
It is worth noting that, given $\cG(z)$, it associates with this function the family of its translations, namely $\cG_n(z) = \cG(z + n d)$. This formalism also allows one to define the transform $G_\beta$ of a discrete variable $G_n$. 

In agreement with Fourier theory, the Gel'fand inverse $\beta$-transform for the product of two $\beta$-based functions is
\begin{align}
\frac{1}{2 \pi} \int^{\pi}_{-\pi} \mathfrak{A}_{\beta} \mathfrak{B}_{\beta}  \, \rme^{-\rmi n \beta d} \, \rmd (\beta d) 
&= \sum_{n'\in \ZZ} \mathfrak{A}_{n'} \mathfrak{B}_{n-n'}  \, ,
\label{e:convol}
\end{align}
for arbitrary functions $\mathfrak{A}$ and $\mathfrak{B}$.

Electromagnetic fields are governed by the Maxwell equations with sources, 
along with boundary conditions of the waveguide (perfect metallic conductor).
We decompose the radiated fields on the relevant function basis in the $\beta$-based representation.
Inside the cell volume $\VV_0$, they satisfy the Helmholtz equations\cite{kuz80,rys09}
\begin{align}
  \nabla \times \bcE^s_{\beta}(\bfr) 
  & = - \rmi \mu_0 \Omega^s_{\beta} \bcH^s_{\beta}(\bfr) \, ,  
  \label{e:Helmo1}  
  \\
  \nabla \times \bcH^s_{\beta}(\bfr) 
  & = \rmi \epsilon_0 \Omega^s_{\beta} \bcE^s_{\beta}(\bfr) \, ,  
  \label{e:Helmo2}
\end{align}
with real eigenvalues $\Omega^s_{\beta}$,
for solenoidal eigenfields ($\nabla \cdot  \bcE^s_{\beta} = 0$, $\nabla \cdot  \bcH^s_{\beta} = 0$)  
satisfying the Floquet and boundary conditions ($\cE$ normal to the perfectly conducting walls, $\cH$ tangent to them). 

We take $\Omega^s_\beta \geq 0$ with no loss of generality (allowing $\beta < 0$ for modes with negative phase velocity, 
with eigenfield $\bcE_{- \beta} = {\bcE_{\beta}}^*$). 
The infinite discrete set of solutions of the Helmholtz equations \eqref{e:Helmo1}-\eqref{e:Helmo2} are labelled with the waveguide band index $s \in \ZZ$
but, in general, there is one dominant ($s=0$) mode of propagation.
The eigenvalues $\Omega^s_{\beta}$ define the dispersion relation (Brillouin diagram) of the structure,  
so that the phase velocity is $v_{\rm ph} (s,\beta) =  \Omega^s_{\beta} / \beta$ 
and the group velocity is $v_{\rm g} (\beta) = \partial_\beta \Omega^s_{\beta}$.
Figure~\ref{f:dispRelImp} displays the dispersion relation
and the interaction impedance (see below) of the 3 m long helix TWT.
For a given wavenumber, the set of electric eigenfields $\bcE^s_{\beta}$ 
is an orthogonal basis of the Hilbert space of divergence-free square-integrable fields on $\VV_0$ 
which have a square-integrable curl and are normal to the perfectly conducting wall of the cell.
Eigenfields $\bcE^s_{\beta},\bcH^s_{\beta}$ generate the time-independent basis fields 
depending only on the structure geometry.

\begin{figure}
\centering
\includegraphics[width=\columnwidth]{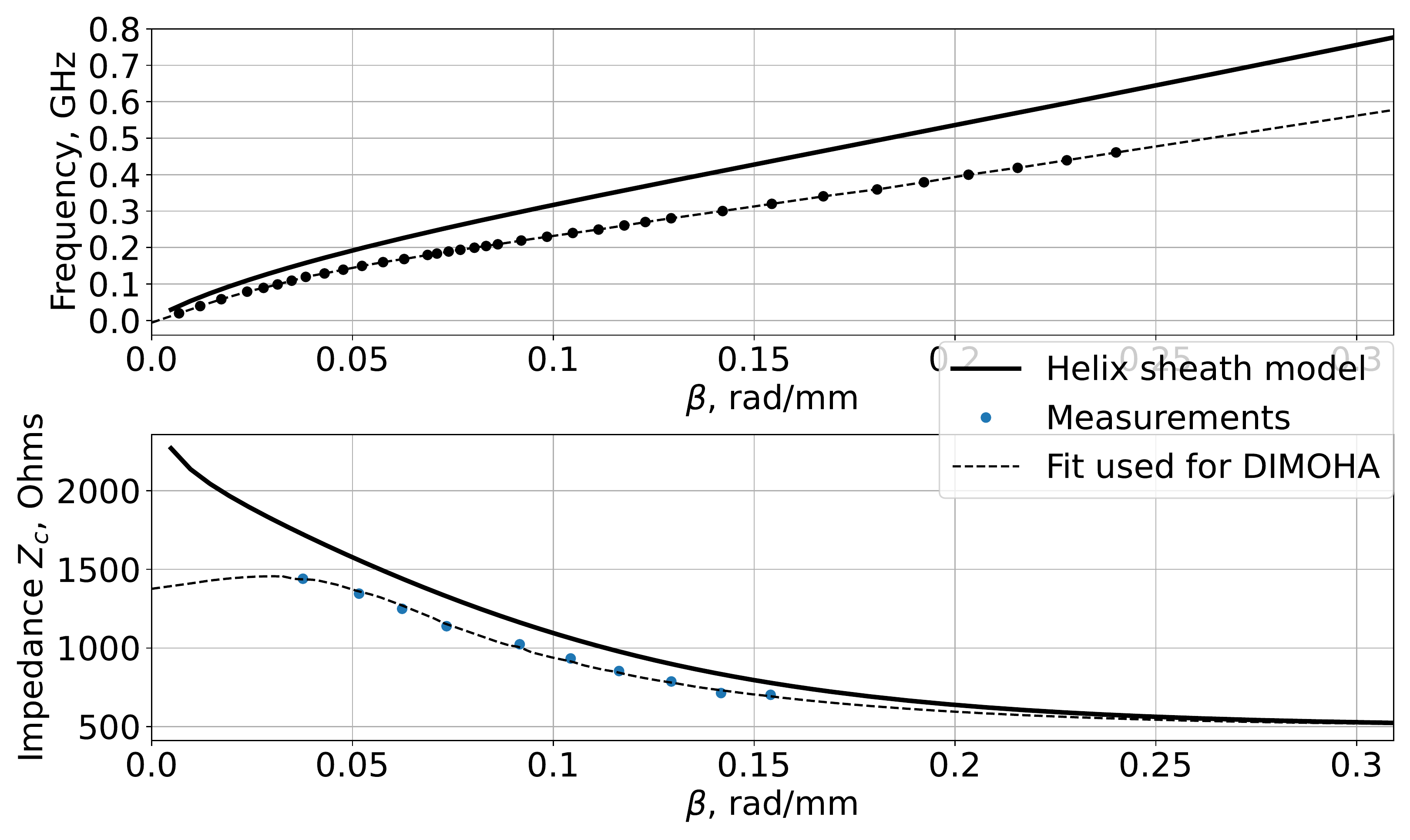} 
\caption{Dispersion relation $\Omega_{\beta} / (2 \pi)$ (upper panel) and coupling impedance $Z_\rmc(\beta)$ (lower panel) 
     as functions of the wavenumber $\beta$ for the dominant ($s=0$) mode of propagation.
     Measurements (dots) and fit (dashed line) from the 3~m long helix TWT from Ref.~\onlinecite{dim78}. 
The continuous lines are from the dispersion relation Eq.~\eqref{e:bondaBessel5} and impedance Eq.~\eqref{e:3DZc}.
Both continuous lines used the sheath helix model from Section~\ref{s:sheathDIMO} with pitch $d = 2.54$~mm and radius $a = 8.06$~mm as for the 3~m long TWT.
The time-domain simulation in Section~\ref{s:validation} is performed at $F=0.22$~GHz with the dashed lines.}
\label{f:dispRelImp}
\end{figure}

\subsection{Field decompositions}
\label{s:FieldsDimo}

With the Gel'fand $\beta$-transform \eqref{e:fourierBeta}, we can re-express the electric field as
\begin{equation}
\mathbb{E}_\beta(\bfr,t) = \sum_{n \in \ZZ} \bfE (\bfr+ n d \, \bfe_z,t)  \, \rme^{ \rmi n \beta d} \, ,
\end{equation}
with the unit vector $\bfe_z$ along the longitudinal axis,
with the property $\mathbb{E}_\beta = \mathbb{E}_{\beta + 2 \pi /d}$,
and the same for the magnetic field $\mathbb{H}_\beta$.
Any divergence-free field $\mathbb{E}_\beta$ (respectively $\mathbb{H}_\beta$) meeting the Floquet condition \eqref{e:Floquet} 
can be decomposed on the eigenfields $\bcE^s_{\beta}(\bfr)$ (respectively $\bcH^s_{\beta} (\bfr)$)
with time-dependent amplitudes $\cV^s_{\beta}(t)$ (respectively $\cI^s_{\beta}(t)$) 
as
\begin{eqnarray}
  \mathbb{E}_\beta(\bfr,t) 
  & = & \sum_s \cV^s_{\beta} (t) \bcE^s_{\beta} (\bfr) \, ,
  \\
  \mathbb{H}_\beta(\bfr,t) 
  & = & \sum_s \rmi \, \cI^s_{\beta} (t) \bcH^s_{\beta} (\bfr) \, .
\end{eqnarray}
Using Eq.~\eqref{e:betainverse}, 
we re-express radiofrequency (RF) electromagnetic (a.k.a.\ circuit) fields 
in the $\beta$-based representation as
\begin{align}
  \bfE(\bfr,t) 
  & = \frac{1}{2 \pi} \sum_{s \in \ZZ} \int^{\pi}_{-\pi} \cV^s_{\beta}(t) \, \bcE^s_{\beta}(\bfr) \, \rmd (\beta d) \, , 
  \label{e:VbetaEbeta}  
  \\
  \bfH(\bfr,t) 
  & = \frac{1}{2 \pi} \sum_{s \in \ZZ} \int^{\pi}_{-\pi} \rmi \, \cI^s_{\beta}(t) \, \bcH^s_{\beta}(\bfr) \, \rmd (\beta d) \, ,
  \label{e:IbetaHbeta}
\end{align}
with time-dependent amplitudes $\cV^s_{\beta}(t)$, $\cI^s_{\beta}(t)$.
The imaginary number ($\rmi$) is for convenience in view of \eqref{e:Helmo1}-\eqref{e:Helmo2}.

Along with this solenoidal field \eqref{e:VbetaEbeta} (circuit part), 
an additional curl-free electric field $-\nabla \phi$ (space charge part) is generated by the presence of charged particles, 
with the scalar potential $\phi$ satisfying the Poisson equation $\Delta \phi = - \rho / \epsilon_0$. 
Our model separates the electric field in the circuit part~\eqref{e:VbetaEbeta} and the space charge part. 
Since the $\beta$-transform commutes with both time and space derivatives, 
Maxwell equations apply to the $\beta$-representation of fields.\cite{and13}
Until Sub-Section~\ref{s:SC}, we omit the space charge of particles.

Thanks to the Gel'fand $\beta$-transform, electromagnetic fields from Eqs~\eqref{e:VbetaEbeta}-\eqref{e:IbetaHbeta}, that satisfy Maxwell's equations, can also be decomposed in the representation per cell as\cite{and13}
\begin{align}
\bfE(\bfr,t) & = \sum_{s,n} \cV^s_{n}(t) \, \bcE^s_{-n}(\bfr) \, , 
\label{e:VsnEsn} \\
\bfH(\bfr,t) & = \rmi \sum_{s,n} \cI^s_{n}(t) \, \bcH^s_{-n}(\bfr) \, , \label{e:IsnHsn}
\end{align}
where $\cV^s_{n}$, $\cI^s_{n}$ are the discrete (cell-based) temporal field amplitudes 
and $\bcE^s_{n},\bcH^s_{n}$ the ``shape'' functions of the fields depending only on the structure geometry.
The sign in $-n$ is a technical convention\cite{and13,min18} analogous to the one in convolutions in agreement with Eq.~\eqref{e:convol}. 

We also express the circuit vector potential as
\begin{equation}
\bfA(\bfr,t) =  \rmi \sum_{s,n} \cI^s_{n}(t)  \bcA^s_{-n}(\bfr) \, , \label{e:potVec}
\end{equation}
from $\nabla \times \bfA = \mu_0 \bfH$ since the slow wave structure is in vacuum 
in presence of purely dielectric materials. 
With the Weyl gauge, the circuit vector potential eigenfield is $\bcA^s_\beta = \rmi \bcE^s_\beta / \Omega^s_\beta$. 

To be able to express the electromagnetic field, one needs to find the time amplitudes and the shape functions.
The time amplitudes $\cV^s_{n}$, $\cI^s_{n}$ depend on the interaction of the field with charged particles.
Therefore, one needs an interaction model. 
In this paper, we propose the multi-particle description to find both amplitudes (see Section~\ref{s:1DNbodyHam}).
One can also use a fluid model in frequency domain as in Ref.~\onlinecite{min19ps}.
The shape functions $\bcE^s_{n}$, $\bcH^s_{n}$ depend on the structure geometry.
In 1D, one can use an approximation\cite{rys09,min18} from the harmonic power to express the projection, on the longitudinal $z$-axis, of shape fields $\bcE^s_{n}$  (see Eq.~\eqref{e:Ebeta1D}).
One of the main goals of this paper is to express eigenfields $\bcE^s_{n},\bcH^s_{n}$ using the helix geometry (see Section~\ref{s:sheathDIMO}). 

Beside the time-domain approach, one can use a frequency-domain version of Kuznetsov's discrete model \cite{rys07,rys07ieee,min19ps,ter17} in small signal regime. 
The validity of the latter was assessed \cite{min19ps} by comparing it with Pierce's four-waves theory.\cite{pie50}


To complete the model, we normalize the eigenfields to\cite{rys09,and13}
\begin{equation}
\Omega^{s_1}_{\beta} \delta_{s_2}^{s_1} = \int_{\VV_{0}} \epsilon_0 \bcE^{s_1}_{\beta} \cdot {\bcE^{s_2}_{\beta}}^* \, \rmd V
  = \int_{\VV_{0}} \mu_0  \bcH^{s_1}_{\beta} \cdot {\bcH^{s_2}_{\beta}}^* \, \rmd V \, ,
  \label{e:normal}
\end{equation}
with $\VV_{0}$ the cell domain. 
Note for later that Eq.~\eqref{e:normal} is stated for a single wavenumber $\beta$.
For convenience,\cite{and13}  this normalisation is chosen equal to the eigenvalue $\Omega^s_{\beta}$ in Helmholtz' equations \eqref{e:Helmo1}-\eqref{e:Helmo2},
which is the angular frequency $\omega = \Omega^s_{\beta}$ in the dispersion relation.
The Kronecker symbol $\delta_{s_2}^{s_1} $ expresses the orthogonality of modes, 
if eigenvalues are non-degenerate ($\Omega_\beta^{s_1} \neq \Omega_\beta^{s_2}$ if $s_1 \neq s_2$), 
whatever the geometry of the cells.
For cylindrical structures, this orthogonality can be extended to the cross product in the expression of the electromagnetic power $\int_{\VV_0} (\bcE^{s_1}_{\beta} \times \bcH^{s_2 *}_{\beta} ) \cdot \bfe_z \, \rmd V =0$, with $s_1 \neq s_2$.
In the rest of the paper, we imply only modes $s_1 = s_2$.

Our choice to have $\Omega^s_\beta \geq 0$ and $\bcE^{s *}_{\beta} = \bcE^{s}_{-\beta}$ 
implies the reciprocity conditions  $\bcH^{s *}_{\beta} = - \bcH^{s}_{-\beta}$,  
$\cV^{s *}_{\beta} = \cV^{s}_{-\beta}$ and  $\cI^{s *}_{\beta} = \cI^{s}_{-\beta}$.
Therefore, $\Omega^s_n$, $\cV^{s}_{n}$, $\cI^{s}_{n}$, $\bcE^{s}_{n}$, $\rmi \bcH^{s}_{n}$, $\rmi \bcA^{s}_{n}$ 
are real-valued.
We also deduce $\cV^{s}_{\beta} =  \int_{\VV_{0}} \epsilon_0 \, \mathbb{E}_\beta \cdot {\bcE^{s *}_{\beta}}  / \Omega^s_\beta \, \rmd V$ and $\cI^{s}_{\beta} =   \int_{\VV_{0}} \mu_0 \, \mathbb{H}_\beta \cdot {\bcH^{s *}_{\beta}} / \Omega^s_\beta\, \rmd V$.

Two of the initial inputs of our model are the (cold) dispersion relation $\Omega^{s}_\beta \in \RR_{\geq 0}$ 
and the interaction (coupling) impedance $Z_{\rm c}(\beta) \in \RR_{\geq 0}$, 
used below to compute the eigenfield functions (see Eq.~\eqref{e:Ebeta1D}). 
Both consider only the wave propagation in the waveguide without the beam.
For 15 cm long TWTs, we use an industrial numerical solver that computes both 
from the wave equations with boundary conditions, taking into account features 
like materials, wire thickness, or dielectric support rods.
For large TWTs, the dispersion relation and the interaction impedance can be directly measured 
(see Fig.~\ref{f:dispRelImp}).

\begin{figure}
\centering
\includegraphics[width=\columnwidth]{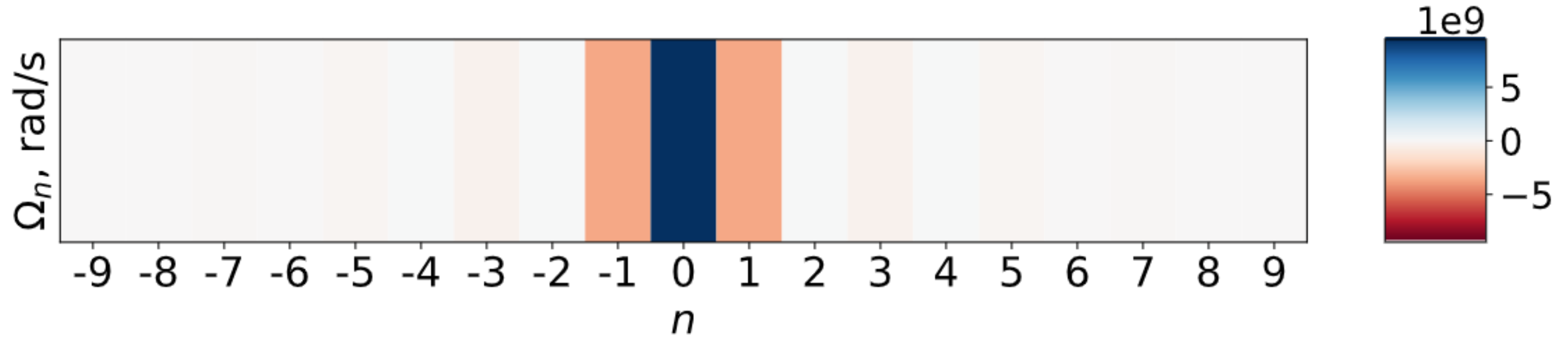}
\caption{Coupling coefficients $\Omega^s_{n}$ \eqref{e:OmegaN} from the transform \eqref{e:betainverse} of the dispersion relation $\Omega^s_{\beta}$. 
We use the dispersion relation of the sheath helix model displayed in Fig.~\ref{f:dispRelImp}. }
\label{f:omegaN}
\end{figure}

\subsection{Wave propagation with the discrete model}
\label{s:WavePropagation}

From our field decomposition \eqref{e:VsnEsn}-\eqref{e:IsnHsn}, normalisation \eqref{e:normal} and Helmholtz' equations \eqref{e:Helmo1}-\eqref{e:Helmo2}, both Maxwell-Amp{\`e}re and Maxwell-Faraday equations can be rewritten.
In representation per cell, 
this yields\cite{and13}
\begin{align}
  \frac{\rmd \cV^s_n}{\rmd t} 
  &= - \sum_{n' \in \ZZ} \Omega^s_{n-n'} \cI^s_{n'}(t)  + \int_{\VV_\ZZ} \bfJ(\bfr,t)  \cdot \rmi \bcA^s_{- n}(\bfr)  \, \rmd V \, , 
  \label{e:max1} 
  \\
  \frac{\rmd \cI^s_n}{\rmd t} 
  &=  \sum_{n' \in \ZZ} \Omega^s_{n-n'} \cV^s_{n'}(t)  \, , 
  \label{e:max2}
\end{align}
where the coupling factors (see the description below)
\begin{equation}
\Omega^s_{n} = \frac{1}{2 \pi} \int^{\pi}_{-\pi} \Omega^s_{\beta} \, \rme^{-\rmi n \beta d} \, \rmd (\beta d)  \label{e:OmegaN}
\end{equation} 
are the Fourier coefficients \eqref{e:betainverse} of the dispersion relation and $\bfJ$ is the current density. Figure~\ref{f:omegaN} displays the coupling coefficients \eqref{e:OmegaN} for the helix geometry. 
The normalisation (\ref{e:normal}) implies \cite{and13} that the source term in (\ref{e:max1}) involves the shape fields 
$\bcA^s_n$,
associated with the vector potential. 
In our approach,\cite{and13,min18} 
we define the coupling coefficients $\Omega^s_{n}$ to be homogeneous to a pulsation, 
so the time amplitudes $\cV^s_{n}$ and $\cI^s_{n}$ have the dimension of the square root of an action, 
but the coupling coefficients can also be chosen to have\cite{rys09} the dimension of an energy 
with dimensionless $\cV^s_{n}$ and $\cI^s_{n}$.

For a source-free system ($\bfJ = {\textbf{0}}$), we see that the discrete model represents the wave propagation 
as an infinite chain of oscillators with $\Omega^s_{n-n'}$ the coupling coefficient between cells $n$ and $n'$.
Those coupling coefficients, which depend on the tube geometry, are computed from the dispersion relation (Brillouin diagram) 
$\Omega^s_{\beta} = \sum_{n}  \Omega^s_{n} \, \rme^{\rmi n \beta d}$.
For practical applications, note that the accuracy of those coupling coefficients depends on the sampling of the dispersion relation.
In particular, the discrete model requires a wider dispersion relation 
(including small and large wavenumbers $\beta$) 
than just the data for the transmission band. 
While this may seem a drawback compared with frequency models (for which a single point in the dispersion relation suffices),
it is an advantage for the discrete model as it will enable broadband modelling (e.g.\ pulse propagation~\cite{ali21}) and include harmonics in nonlinear regimes.
This advantage was used\cite{kuz84,bes00,kuz04} to investigate instabilities\cite{mil94,hun15,ant18} at cutoff (near the edge of the transmission band) of TWTs where frequency models have difficulties to describe the situation.
Originally, the discrete model was developed to investigate those regimes, especially for coupled cavity TWTs.

The first major advantage of the discrete model as a reduction model is that, 
when considering only the dominant mode of propagation $s=0$, 
the electromagnetic field~\eqref{e:VsnEsn}-\eqref{e:IsnHsn} requires only 
$n_{\rm max}$ couples of the time variables $\cV_{n}, \cI_{n}$, 
viz.\ $n_{\rm max}$ DOFs for $n_{\rm max}$ cells. 
Time-independent functions $\bcE_{n},\bcH_{n}$ and coupling coefficients $\Omega^s_{n}$ depend only on the structure geometry. 
To give an order of magnitude, 
space TWTs ($\sim 15$~cm) with helical SWSs to amplify waves in the GHz range 
are $n_{\rm max} \approx 200$~periods long, 
therefore the field model involves only 200 DOFs $(\cV_{n},\cI_{n})$. 
Sub-THz tubes, that mostly use coupled cavities or folded wave\-guides\cite{and20ted}, 
are $n_{\rm max} \approx 20$--$100$ periods long ($\sim 2$ to 10 cm).
The San Diego 3 m long TWT, operating at 200 MHz, is composed of $n_{\rm max} \approx 1000$ periods, while the Marseille 4 m long TWT, operating at 30 MHz, is composed of $n_{\rm max} \approx 4000$ periods.

In principle, the range of coupling coefficients $\Omega^s_{n-n'}$ should be infinite. 
But in the discrete model, the dispersion relations $\omega (\beta) = \Omega^s_{\beta}$ for each mode $s$ 
are decomposed in a Fourier basis on $[ - \pi / d , \pi / d ]$, and in TWTs, 
those dispersion relations are similar to cosine functions (for large phase $\beta d$). 
Therefore, only the first few coefficients of the Fourier series are non-negligible ($\lim_{n \rightarrow \infty} \Omega^s_{n} =0$), 
and the coupling between cells involves mainly nearby cells, 
$\Omega^s_{\beta} \simeq \sum^{N_{\rm ph}}_{n = -N_{\rm ph}}  \Omega^s_{n} \, \rme^{\rmi n \beta d} =  \Omega^s_{0} + \sum^{N_{\rm ph}}_{n =1} 2 \Omega^s_{n} \cos (n \beta d)$.
The range of the coupling was estimated \cite{ber11} to $N_{\rm ph} \in [5,20]$ for helix TWTs 
and $N_{\rm ph} \approx 1$ for folded wave\-guide TWTs\cite{min19ted,and20ted}. 
Figure~\ref{f:omegaN} shows that the coupling terms, for a helix waveguide, are negligible for $n > 5$. 
To limit boundary effects, it is reasonable to limit the range of coupling taken.
Thanks to this finite range, we can investigate real devices 
because the boundary at the tube ends does not affect the propagation in the middle.
This also reduces the weight of calculations for the field propagation.

In actual devices, the wave propagation should take into account losses, like skin effects or dielectric losses.
These can be modelled by rewriting eq.~\eqref{e:max1}, with $\bfJ=0$, as
\begin{equation}
  \frac{\rmd \cV^s_n}{\rmd t} 
  = - \sum_{n' \in \ZZ} \Omega^s_{n-n'} \cI^s_{n'}(t) - \alpha^s_n  {\cV}^s_{n} \, , 
  \label{e:attenuationevolution}
\end{equation}
with an attenuation rate $\alpha$ (per time unit) derived from the conductivity.

To avoid the total wave reflection  from an abrupt stop of the harmonic chain at the tube edges (a boundary effect), 
we use a perfectly matched layer (PML).
We add attenuation coefficients $\alpha_n$ that slowly increase when moving away from the physical part.
With the adjustment of those coefficients, 
we can generate a small edge reflection (a.k.a.~voltage standing-wave ratio, VSWR). 
This reflection occurs in real tubes from adaptation defects between the SWS and its wave output port.

The approach presented in this paper rests on a periodic structure model. 
However, industrial TWTs are designed with pitch $d$ and radius $a$ variations, called helix tapering.
This implies that the dispersion relation changes with the cell.
The passage from a cell with a given dispersion relation to another one with a different dispersion relation 
generates defects in the wave propagation (reflections and oscillations).
To take the pitch and radius tapering into account, 
we first compute the coupling coefficients for each cell $n_1$ 
to build the arrays $\Omega^{s,(n_1)}_{n_2} = \Omega^{s,(n_1)}_{- n_2}$ for the ranges 
$- N_{\mathrm{ph}} \leq n_2 \leq N_{\mathrm{ph}}$ (possibly expanded with zeros for $| n_2 | > N_{\mathrm{ph}}$). 
We ensure the symmetry of coupling coefficients between cells $n$ and $n'$ in Eqs ~\eqref{e:max1}-\eqref{e:max2}
by replacing $\Omega^s_{n - n'}$ with the averages 
$\bar \Omega^s_{n, n'} = \bar \Omega^s_{n', n} = (\Omega^{s,(n)}_{n - n'}  + \Omega^{s,(n')}_{n -n'} ) /2$.
 This may violate slightly the discrete model derivation, 
but the variation of the dispersion relation from a cell to its $N_{\mathrm{ph}}$ neighbours is small enough, 
in industrial TWTs, for the adiabatic approximation to hold. 
However, the 3 m long TWT has no tapering and keeps the same dispersion relation all along the tube.

\section{One-dimensional multi-particle self-consistent Hamiltonian model}
\label{s:1DNbodyHam}

\subsection{Section synthesis}

In this section, we combine the discrete model for fields with a 1D multi-particle model for the particle dynamics.
The discrete model is used to construct a total self-consistent Hamiltonian that allows us to express the evolution equations of the time-dependent variables of the system.
From it, we formulate a numerical integrator for the total dynamics.
Space charge effects are taken into account using Rowe's disk model.

\subsection{One-dimensional eigenfields and shape factors}
\label{s:1Dfields}

An important advantage of the discrete model is that it is not limited to a particular geometry 
but can be applied to any structures periodic along the longitudinal $z$-axis with $\bfe_z$ its unit vector. 
The discrete model can be projected onto the longitudinal $z$-axis in one dimension (1D). 
We can do so because electric fields near the $z$-axis are almost longitudinal in TWTs 
and because we have collimated electron beams with small radius.

\begin{figure}
\centering
\includegraphics[width=\columnwidth]{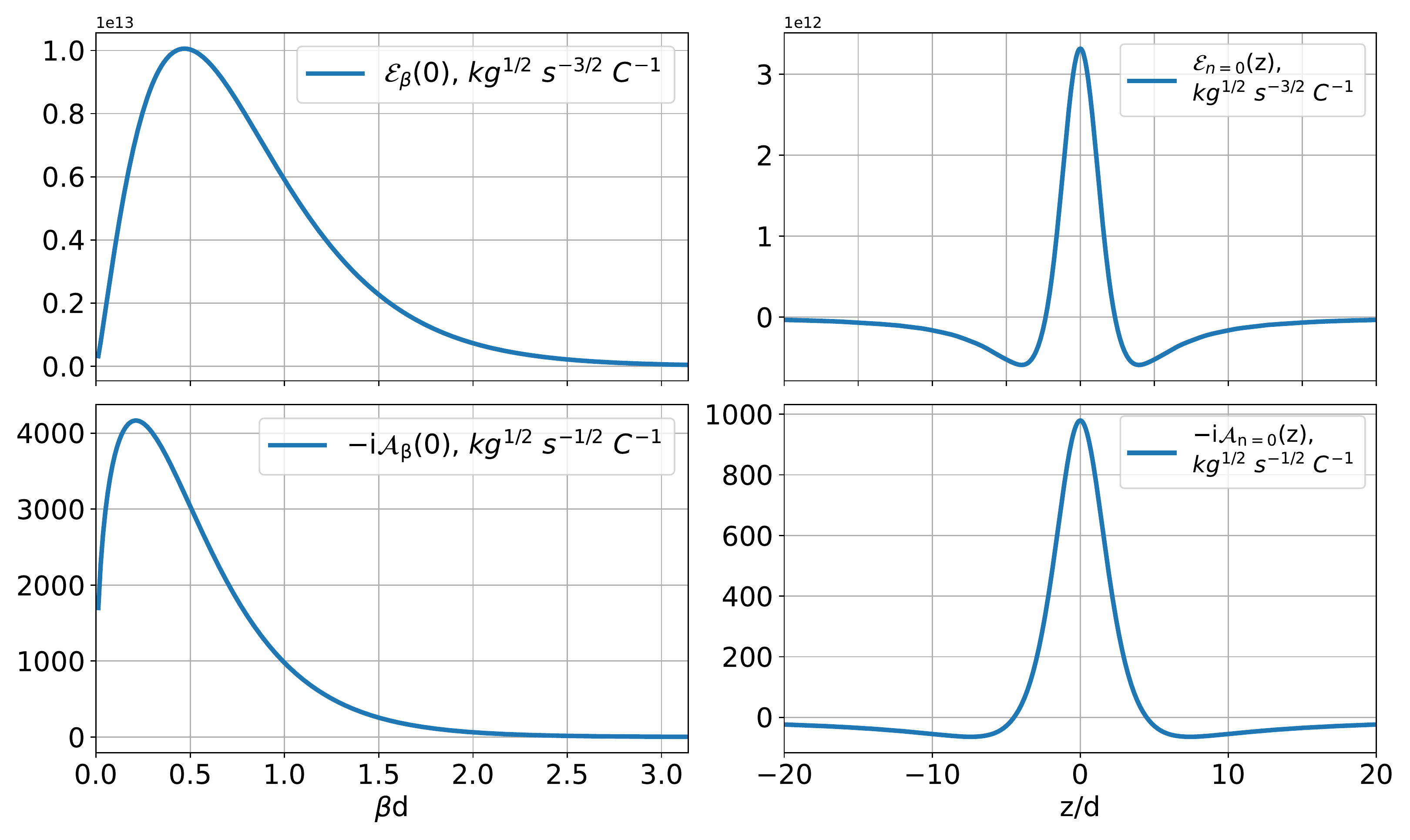}
\caption{
Amplitude of eigenfields (left column) and shape functions (right column), for the electric field (upper row) and the vector potential (lower row), from Eqs~\eqref{e:Ebeta1D}-\eqref{e:Anz}.
All panels use the sheath helix dispersion relation and sheath helix interaction impedance from Fig.~\ref{f:dispRelImp} as function of the phase per pitch $\beta d$. }
\label{f:EbetaEn}
\end{figure}

The model is projected in 1D along the $z$ axis. The static magnetic field that confines the beam is neglected. 
We note $\bcE^s_{\beta} = \cE^s_{z, \beta}(z) \, \bfe_z$ the electric eigenfield function as the inverse Fourier transform (viz.\ Gel'fand) of $\bcE^s_{n}$. 
The fields are assumed longitudinal (which is a good approximation on the $z$-axis).
From the harmonic power \eqref{e:powerHarmo}, a 1D eigenfield for any geometry 
was already obtained,\cite{rys07,rys07ieee,min18} 
\begin{equation}
  \cE^s_{z, \beta} (z) 
  = \rme^{-\rmi \beta z}  \beta \, \sqrt{\frac{1}{d} \Omega^s_\beta Z_{\rm c}^s (\beta) v_{\rm g}^s (\beta)} \, , 
  \label{e:Ebeta1D}
\end{equation}
for $0 < \beta d \leq \pi$ in terms of cold-field (beamless) parameters: 
the wavenumber (propagation constant) $\beta$, 
angular frequency $\Omega^s_\beta$ given by the dispersion relation, 
group velocity $v_{\rm g}^s$ and interaction (coupling) impedance $Z_{\rm c}^s$. 
With the Weyl gauge, the projected vector potential eigenfield is 
\begin{equation}
  \cA^s_{z, \beta} (z) 
  = \frac{\rmi \cE^s_{z, \beta} (z)}{\Omega^s_\beta} \, . 
  \label{e:AsbetaEQiEsbeta}
\end{equation}
The left column of Fig.~\ref{f:EbetaEn} displays the electric field (upper row) and vector potential (lower row) eigenfield amplitudes given by the helix geometry.
Using the Gel'fand inverse $\beta$-transform \eqref{e:betainverse} on Eqs \eqref{e:Ebeta1D} and \eqref{e:AsbetaEQiEsbeta}, we obtain the longitudinal values of the electric shape function
\begin{equation}
\cE_{z,n} (z) 
  = \frac{1}{2 \pi} \int^{\pi}_{-\pi} \cE^s_{z, \beta} (z)  \, \rme^{-\rmi n \beta d} \, \rmd (\beta d) \, , 
  \label{e:Enz}
\end{equation}
needed to express the 1D real electric field \eqref{e:VsnEsn},
and the vector potential shape function
\begin{equation}
\cA_{z,n} (z) 
  = \frac{1}{2 \pi} \int^{\pi}_{-\pi} \cA^s_{z, \beta} (z)  \, \rme^{-\rmi n \beta d} \, \rmd (\beta d) \, ,
  \label{e:Anz}
\end{equation}
needed to express the 1D real vector potential \eqref{e:potVec}. 
The shape functions \eqref{e:Enz} and \eqref{e:Anz} depend only 
on the structure geometry contained in the dispersion relation and interaction impedance. 
The right panels of Fig.~\ref{f:EbetaEn} display them for the helix geometry.
To complete the model and to express the electric field \eqref{e:VsnEsn} and vector potential \eqref{e:potVec}, 
we only need to find the values of the time amplitudes $ {\cV}^s_{n}$ and ${\cI}^s_{n}$.

As displayed in Fig.~\ref{f:EbetaEn}, the electric field and vector potential shape functions 
$\cE_{z,n} (z)$ and $\cA_{z,n} (z)$ are smooth functions. 
This feature is crucial for us since the vector potential is at the heart of the coupling between waves and particles. 
This means that, for two given close positions, the coupling is almost identical. 
Therefore, instead of computing with each individual electron, 
we aggregate them to form so-called macro-electrons. 

Each macro-electron is treated as a single point particle with a position $q_k$ and velocity $\dot{q}_k$.
Their electric charge is $e = - |I_0| \delta / v_0$, with $I_0$ the cathode current
and $\delta$ the initial distance between two macro-electrons.
The initial beam velocity is $v_0 = c [1 - (1 - e_{\rm 0} V_0/(m_{\rm 0} c^2))^{-2}]^{1/2}$, 
with $V_0$ the cathode potential, $e_{\rm 0} < 0$ the electron elementary charge, $m_{\rm 0}$ the electron mass, 
when assuming the initial beam to be a line of equally spaced particles.
Even if the relativistic correction is small, we choose to build the model and algorithm with relativistic dynamics.
The macro-electron mass is $m = m_{\rm 0} e / e_{\rm 0}$.
The spacing $\delta$ is a numerical parameter that is minimized to ensure the convergence of simulations.

\subsection{The wave-particle dynamics}
\label{s:wavepartdynamics}

Having obtained the eigenfields for the tube geometry, 
we now want to express the time evolution of amplitudes $ {\cV}^s_{n}$ and ${\cI}^s_{n}$ 
from the wave-particle interaction.
We combine the discrete model with the Hamiltonian approach.
Inserting the decomposition \eqref{e:VsnEsn}-\eqref{e:IsnHsn} 
in the classical electromagnetic Hamiltonian\cite{jac99} (energy) 
$\frH_{\rm em} =  \int_{\VV_\ZZ} (\epsilon_0 |\bfE|^2 +  \mu_0 |\bfH|^2)/2 \, \rmd V$ leads to \cite{and13,min18}
\begin{equation}
  \frH_{\rm em} = \sum_{s \in \ZZ}  \sum_{n_1,n_2}  \frac{1}{2}
          \left( {\cV}^s_{n_1} {\cV}^s_{n_2} + {\cI}^s_{n_1} {\cI}^s_{n_2} \right) \Omega^s_{n_1-n_2}  \, , \label{e:Hem}
\end{equation}
where the coupling coefficients $\Omega^s_{n_1-n_2}$ between cells $n_1$ and $n_2$
have a limited range ($| \Omega^s_{n} | / \Omega^s_0 \approx 0$ for $|n| > N_{\rm ph}$). 
For helix simulations, $N_{\rm ph} = 5$ is enough.
The time amplitudes become the canonical variables of the field with the conjugate momenta ${\cV}^s_{n}$ and  generalized coordinates ${\cI}^s_{n}$. The normalisation \eqref{e:normal} ensures that they satisfy the Poisson brackets $\{ {\cV}^s_{n_1}, {\cV}^{s'}_{n_2} \} = 0$, $\{ {\cI}^s_{n_1}, {\cI}^{s'}_{n_2} \} = 0$ and $\{ {\cV}^s_{n_1}, {\cI}^{s'}_{n_2} \} = \delta^{n_1}_{n_2} \delta^{s}_{s'}$.

Note that, instead of using the cartesian canonical variables, 
we can use a ``pseudo'' angle-action approach taking the complex-valued time amplitudes
$\mathcal{J}^s_{n} (t) = (\cV^s_{n} (t) + \rmi \cI^s_{n}(t)) / \sqrt{2}$
with $|\mathcal{J}^s_{n}|$ being the canonical ``action'' 
and $\theta^s_n = {\rm arg}\, \mathcal{J}^s_{n}$ its conjugate ``angle''. 
With these variables, the electromagnetic Hamiltonian \eqref{e:Hem} becomes\cite{min18}
  $\frH_{\rm em} = \sum_s  \sum_{n_1,n_2} {\mathcal{J}}^s_{n_1} \, {\mathcal{J}}^{s*}_{n_2} \, \Omega^s_{n_1-n_2}$.
Since the differences $\theta^s_{n_1} - \theta^s_{n_2}$ appear in $\frH$, 
this hamiltonian is however not in pure ``action'' form.

The particle dynamics is governed by the Lorentz force, generated by the relativistic Hamiltonian\cite{jac99} $\frH_{\rm el} = \sum_k m c^2 (\gamma_k - 1)$
for macro-electrons,
with the particle velocities $v_k = \dot{q}_k$, the vector potential $A_z$ on the $z$-axis from the projection of Eq.~\eqref{e:potVec}, 
the particle canonical momentum $p_k = \gamma_k m \dot{q}_k + e A_z(q_k)$ 
and the Lorentz factor $\gamma_k =  ( 1 - |\dot{q}_k|^2 / c^2 )^{-1/2}$.
Note that $\frH_{\rm el}$ depends on the magnetic coefficients $\cI_n$ (via $A_z$) 
but not on the electric variables $\cV_n$.

\subsection{Space charge model}
\label{s:SC}

The above description of the discrete model is limited to (divergence-free) solenoidal electromagnetic waves.
It misses the Coulomb field responsible for the repulsion between electrons. 
The electric field \eqref{e:VsnEsn} is only the circuit part, 
and now we add the space charge field $\bfE_{\rm sc}(\bfr,t) = - \nabla \phi_{\rm sc} (\bfr,t)$, 
with $\phi_{\rm sc}$ the scalar potential of the beam space charge,
which satisfies Poisson's equation $\Delta \phi = - \rho / \epsilon_0$. 
This adds to the system the space charge Hamiltonian $ \frH_{\rm sc} =  \sum^{N_{\rm e}}_{k' \neq k} e \phi_{\rm sc} ( q_{k}-q_{k'}) / 2 $,
for $N_{\rm e}$ macro-electrons, with $e < 0$ their macro-charge.

Since we are working in 1D, we choose Rowe's disk space-charge model\cite{row65} 
where the beam is described as a succession of charged disks repelling each other.
For $N_{\rm e}$ macro-electrons (indexed by $k$) with positions $q_k$, 
the 1D on-axis space-charge field 
$E_{\rm sc}(q_{k}) = - \sum^{N_{\rm e}}_{k' \neq k} \partial_{z} \phi_{\rm sc} \left( q_{k}-q_{k'} \right)$ 
reads \cite{row65,ber11}
\begin{equation}
  E_{\rm sc}(q_{k}) 
  = \sum^{N_{\rm e}}_{k' \neq k} Q \exp \left( - \left| \frac{q_{k}-q_{k'}}{b/2} \right| \right) {\rm sgn}  \left( \frac{q_{k}-q_{k'}}{b/2} \right) \, ,   
  \label{e:EscRowe}
\end{equation}
with $Q = e / (2 \pi \epsilon_0 b^2)$, the charge $e<0$ and the radius $b$ of the cylindrical beam. Effects of the helix radius are small enough to be neglected.\cite{row65}

Though space charge influences the particle dynamics and, hence, wave amplification, 
its effect is marginal\cite{pie50,gil11} compared with that of the amplified modes. 
Yet, it cannot be neglected in the physics, as it underlies the existence of the two beam modes, 
one of which (the ``slow'' one) interacting with the RF carrier wave. 
Comparison with experiment shows that this model \eqref{e:EscRowe} is realistic. 
While this space charge model can be improved, like with a Poisson solver as in PIC codes, it also enables us to drastically reduce the computation time. Since the charges and masses of our macro-electrons are constant, we pre-compute the values of $E_{\rm sc}$ at each position on a mesh. 
After each displacement of the macro-electrons, the space charge force is applied on them with the nearest grid point approximation.

\subsection{Time evolution of the model}

From the total Hamiltonian and the attenuation model \eqref{e:attenuationevolution}, 
the evolution equations are  
\begin{align}
  \dot{\cV}^s_{n} &= -  \derivep{\frH_{\rm tot}}{\cI^s_{n}} - \alpha^s_n  {\cV}^s_{n}   \nonumber \\
  &= - \sum_{n' \in \ZZ} \bar \Omega^s_{n, n'} \cI^s_{n'}  
        + \sum^{N_{\rm e}}_{k=1}  \rmi e \,  \dot{q}_k \, \cA^s_{z, - n} (q_k)  - \alpha^s_n  {\cV}^s_{n}  
  \, ,  \label{e:evo1} \\
  \dot{\cI}^s_{n} &=  \derivep{\frH_{\rm tot}}{\cV^s_{n}} 
  =  \sum_{n' \in \ZZ} \bar \Omega^s_{n, n'} \cV^s_{n'}  
  \, , \label{e:evo2} \\
  \dot{p}_k  &= -  \derivep{\frH_{\rm tot}}{q_k} 
    = e \dot{q}_k \derivep{A_z}{z}(q_k) + e E_{\rm sc}(q_k) 
    \, ,  \label{e:evo4} \\
 \dot{q}_k  &= \derivep{\frH_{\rm tot}}{p_k} = \frac{1}{\gamma_k m} \left( p_k - e A_z(q_k) \right) \, ,  \label{e:evo3} 
\end{align}
for our 4 time-dependent variables.
Those equations are solved using our algorithm \textsc{dimoha}.\cite{min19ted}
We define the matrix
\begin{equation}
  \cM^s 
  = \begin{pmatrix}
        - \alpha^{s}            &    - \bar \Omega^{s} \\
        \bar \Omega^{s}    &            0
     \end{pmatrix}  \, ,
\end{equation}
with four square blocks: $ \alpha^{s}$ is diagonal containing the losses $\alpha^{s}_n$,
and the entries of $\bar \Omega^{s}$ are the coupling coefficients $\bar \Omega^{s}_{n, n'}$.
For relativistic particles (indexed by $k$),
the 1D numerical symplectic order 2 scheme\cite{hai10} from $t$ to $t+h$ can be 
\begin{align}
&\begin{pmatrix}
\cV^s_n |_{t'} \\
\cI^s_n |_{t+h}
\end{pmatrix}  
  = \exp_{\rm M}(\cM^s  h) \begin{pmatrix}
\cV^s_n |_{t -\frac{h}{2}} \\
\cI^s_n |_{t}
\end{pmatrix} 
\, , \label{e:integ1} \\
&q_k |_{t +\frac{h}{2}} 
  = q_k |_{t - \frac{h}{2}} + h \, v_k |_{t} 
  \, ,  \label{e:integ2}\\
&\gamma _k  |_{t+h} \, v_k |_{t+h} 
  = \gamma _k  |_{t} \, v_k |_{t} + h \frac{e}{m} E_{\mathrm{sc}} (q_k |_{t + \frac{h}{2}})  \nonumber \\
  & \qquad \qquad 
    - \frac{e}{m}  \sum_{n \in \ZZ} \rmi \cA^s_{-n} (q_k |_{t + \frac{h}{2}}) \Big( \cI^s_n |_{t+h} - \cI^s_n |_{t} \Big) 
    \, ,   \label{e:integ3} \\
&\cV^s_n |_{t +\frac{h}{2}} = \cV^s_n |_{t'} + e \sum^{N_{\rm e}}_{k=1} \int^{q_k |_{t+\frac{h}{2}}}_{q_k |_{t-\frac{h}{2}}}  \rmi \cA^s_{-n} (q') \, \rmd q'  
\, , \label{e:integ4}
\end{align}
with $h$ the time step and $\exp_{\rm M}$ the matrix exponential.
The initial ($t=0$) values $\cI^s_n |_{0}, \cV^s_n |_{0}, q_k  |_{0},  v_k  |_{0}$, 
for a single mode of propagation $s=0$, are given.
In continuous waveform (CW) regime, the wave is excited at the frequency $F$ 
by adding a drive $\cV^s_{n=1}(t) = U \cos (2 \pi F t)$, with $U$ depending on the input power.

Using the Hamiltonian formalism is important because it enforces built-in conservation properties. 
From this formalism, we construct the symplectic integrator \eqref{e:integ1}-\eqref{e:integ4}. 
Symplectic schemes\cite{hai10} allow to increase the time step $h$ during simulations, 
thus reducing the computation time.
Thanks to the link between the Lagrangian and Hamiltonian formalisms, we can easily express the conservation of total momentum.\cite{min18}

Equations \eqref{e:evo1}-\eqref{e:evo3} describe our dynamics. 
There are several differences between our method and the PIC approach. Our model is written from a Hamiltonian formulation satisfying the Maxwell-Lorentz equations and not from a kinetic description (like the Vlasovian one). Within the Hamiltonian formalism, we describe the dynamics of point particles. Thanks to the discrete model, the wave-particle coupling is performed directly at the particle position, so there is no particle or field weighting like in PIC codes.
However, the use of macro-particles and the equations \eqref{e:evo4}-\eqref{e:evo3} of the particle mover is similar to some PIC codes.
But we do not use distribution functions.
In addition, the complex form of the field propagation and the 3D effects of the space charge 
often require using PIC codes in 3D (or, at least, 2D), while \textsc{dimoha} is built in 1D 
because the discrete model allows projecting fields on the longitudinal direction.

Nevertheless, it is possible to combine the field propagation from the discrete model with a kinetic or even a fluid description of the beam.
In Ref.~\onlinecite{min19ps}, we combined the discrete model with a fluid model in small signal regime.
It is also possible to use a Vlasovian approach to build a PIC code to describe the particle dynamics and to combine it with the field decomposition of the discrete model.
For 1D simulations in the time domain, we suspect that PIC models cannot be faster than our multi-particle discrete model because of our drastic reduction of the number of particles.
At best, the number of variables for the PIC should get close to our reduction, 
but thanks to conservation properties from the Hamiltonian formalism, our model is robust.
However, to build a 2D or 3D version of our multi-particle dynamics \eqref{e:evo1}-\eqref{e:evo3}, the approximation of the space charge of Sub-Section~\ref{s:SC} with the disk model must be changed for a more complete model, for instance, taking into account the external magnetic field used to confine the beam. 
This will add several parameters and we are not sure that advantages of our description can remain practical due to the additional number of DOFs.
But before considering the 2D or 3D propagation of particles, we must first build a 2D or 3D version of the discrete model to express spatial basis functions $\bcE^s_{n}(\bfr)$ and $\bcH^s_{n}(\bfr)$. 
We discuss a helix theory in Section~\ref{s:sheathDIMO}.

\begin{figure}
\centering
\includegraphics[width=\columnwidth]{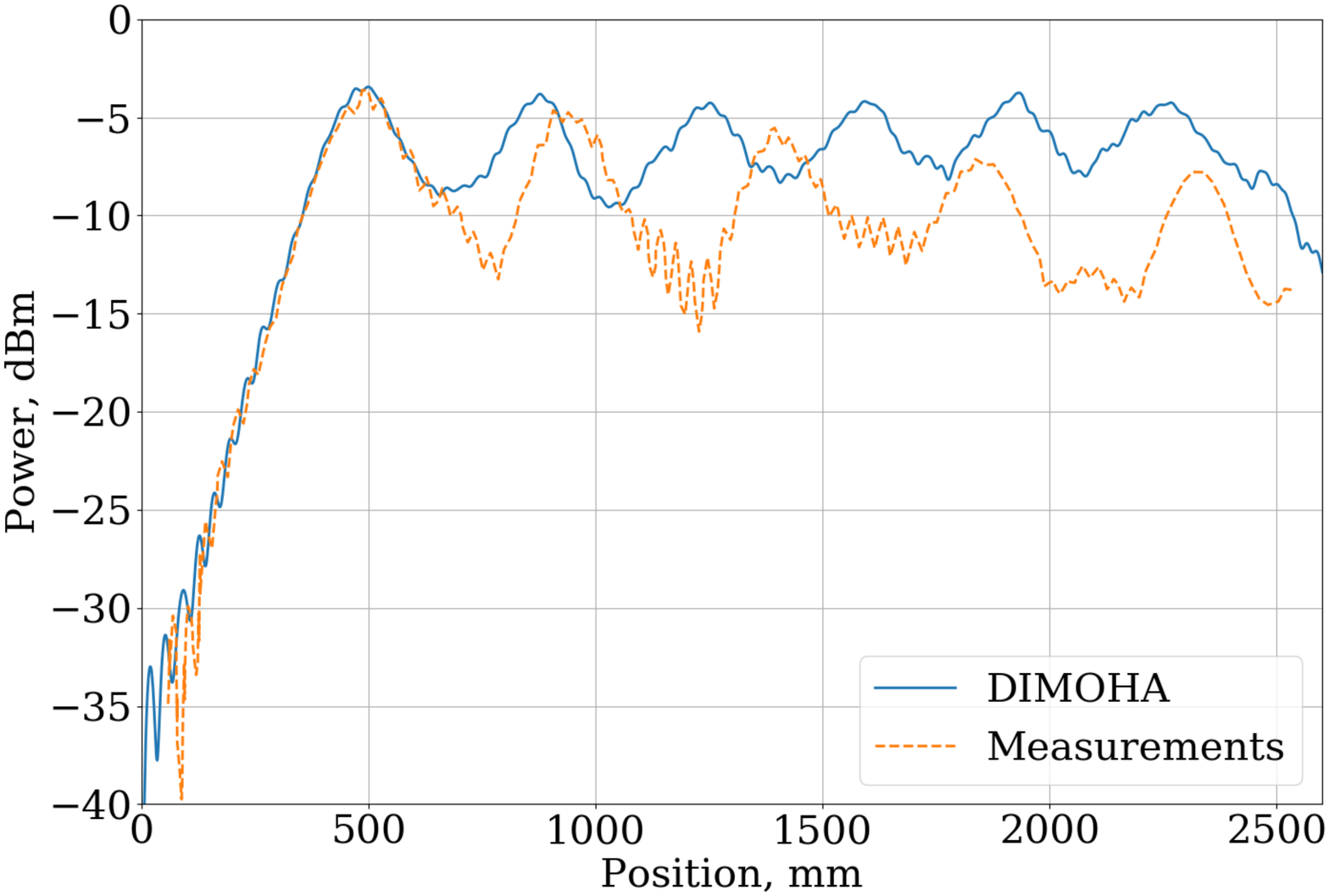} 
\caption{Time average RF power at $F = 220$ MHz as function of the longitudinal position inside a 3 m long TWT\cite{dim78} at time $t = 0.7$ \SI{}{\micro\second}.
The power is expressed in dBm from watts as $P [{\rm dBm}] = 10 \log_{10} (P [{\rm W}] \cdot 10^{3})$.
Simulations are performed using the measured dispersion relation and interaction impedance of the device.}
\label{f:dimonte}
\end{figure}

\begin{figure*}
\centering
\includegraphics[width=\textwidth]{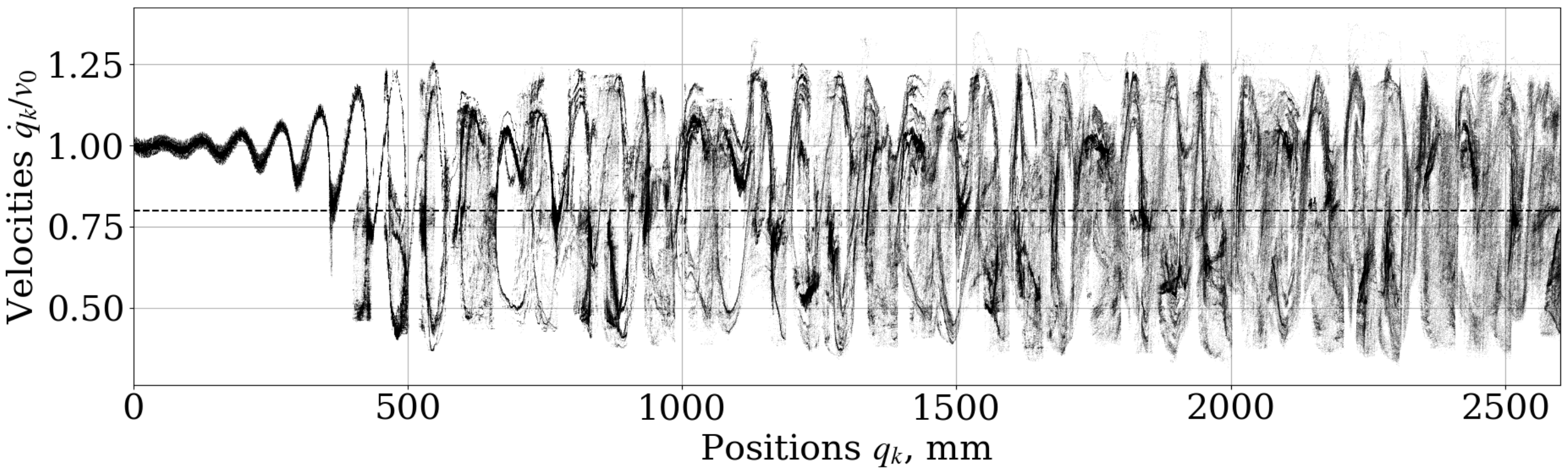} 
\caption{Macro-electrons velocities ratio $\dot{q}_k / v_0$ (black dots) 
     inside a 3~m long TWT\cite{dim78} at time $t = 0.7$~\SI{}{\micro\second} simulated 
     with 4~million particles using \textsc{dimoha} (same simulation as for Fig~\ref{f:dimonte}).
     The dashed line ($v_{\rm ph} / v_0 = 0.8$) corresponds to the phase velocity. 
     Electrons slower than their initial speed ($\dot{q}_k < v_0$) 
     have lost part of their momenta to the benefit of the wave.}
\label{f:dimonteelec}
\end{figure*}

\section{Simulations of a 3-meter long traveling-wave tube}
\label{s:validation} 

\subsection{Section synthesis}

The numerical integrator of the previous section was used to develop \textsc{dimoha}, a code for TWT simulations. 
In this section, \textsc{dimoha} is adapted and validated with an experimental 3 meter long TWT.
Thanks to our reduction model, simulations are fast. Others advantages of \textsc{dimoha} are presented.

\subsection{Simulations}

Equations \eqref{e:evo1}-\eqref{e:evo3} are solved in 1D using our algorithm \textsc{dimoha} described and validated in Ref.~\onlinecite{min19ted} for centimeters long TWTs.
\textsc{dimoha} bears several advantages: 
i)~it can be used to simulate waves with arbitrary waveform and not just the field envelope, 
including multi-carriers (wave amplitudes for multiple frequencies) and digital modulation for telecommunication signals; 
ii)~it is fast (see below), especially compared to PIC codes; 
iii)~it takes into account any periodic structures like helix, coupled-cavity of folded waveguide TWTs; 
iv)~it reproduces the harmonic generation, oscillations, reflection and distortion phenomena; 
v)~it can simulate nonlinear wave-particle effects like trapping or intermodulations.

Computational runs with \textsc{dimoha} are fast thanks to the combination of: 
i)~the DOF reduction from the discrete model; 
ii)~the 1D projection on the $z$-axis allowed by the discrete model; 
iii)~the use of heavy macro-electrons allowed from the smooth shape functions; 
iv)~the fast space charge representation; 
v)~the symplectic integrator \eqref{e:integ1}-\eqref{e:integ4} allowing to increase the time step $h$; 
vi)~ the algorithm CPU parallelisation using message passing techniques. 
Simulations of a 17~cm TWT during 5~ns, displayed on Fig.~\ref{f:newhor}, are performed in only 7~min on a small computer. 
\textsc{dimoha} is about 500 times faster than alternative PIC codes for folded waveguide TWTs for the same results.\cite{thales}

Since then, we are investigating two huge helix TWTs, one 3 m long TWT\cite{dim77,dim78,dim82} and one 4 m long TWT\cite{des20} both used for plasma research, especially in nonlinear regime.
There are no differences in the physical model used to simulate centimeters and meters long TWTs. They are all modelled with Eqs \eqref{e:evo1}-\eqref{e:evo3}.
However, we perform several numerical modifications of \textsc{dimoha} to be able to use it in large scale simulation, including parallelisation and memory management.
We also improved our management of the VSWR using the PML method to adjust the attenuation coefficients $\alpha_n$. 
This adjustment is empirical until we find the correct VSWR matching the measure.

Here, we use \textsc{dimoha} to simulate the San Diego 3~m long TWT.\cite{dim78} 
Its dispersion relation and interaction impedance are displayed in Fig.~\ref{f:dispRelImp}. 
We are only considering the main mode of propagation $s = 0$.
The dispersion relation or impedance of the sheath helix model, 
presented in Section~\ref{s:sheathDIMO}, are not used in this section.
The field shape functions are computed using the eigenfields \eqref{e:Ebeta1D} 
with the measured dispersion relation and impedance.

The wave frequency is $F = 220$~MHz.
The experimental helix pitch is $d = 2.54$ mm, but we lump the cells by groups of four, 
with total pitch $d = 10.16$ mm, to have $n_{\rm cell} = 332$ cells in total. 
This means, we only have 332 amplitudes $\cV_n$ and $\cI_n$ (so 332 DOFs) 
representing the whole electromagnetic field. 
42 cells on each side of the simulations (not displayed in the figures) are used for the PML method. 
The range of coupling is $N_{\rm ph} = 15$. 
One reason we group cells together is because 
the wavelength is quite large at the frequency used ($220$ MHz), 
limiting the sampling of the dispersion relation and the impedance. 
By increasing the pitch, we also increase the phase per pitch, thereby improving our sampling,
especially in view of the limited range of wavenumbers $\beta$ over which it has been measured 
(see Fig.~\ref{f:dispRelImp}).

The initial beam is set with cathode voltage $V_0 = 1$~kV, cathode current $I_0 = 30$~mA, 
beam radius $b = 6$ mm 
and with an initial spacing $\delta = 1$~\SI{}{\micro\meter} between macro-electrons. 
Therefore, there are about $N_{\rm e} = 4$~million particles in our simulations.
The space charge field \eqref{e:EscRowe} is pre-computed on $n_{\rm os}  n_{\rm cell} = 3320$ fixed points 
with oversampling $n_{\rm os} = 10$.
The total simulation time is $t_{\rm max} = 0.7$~\SI{}{\micro\second} with a time step $h = 8.88$~ps. 
The total time is chosen long enough to ensure the simulation reaches a steady state.
The values of $n_{\rm os}, \delta$ and  $h$ are numerical parameters 
chosen to ensure the numerical convergence of simulations.
The computation time is 9.8~h using the 32 processors 
of an Intel\textsuperscript{\textregistered} Xeon\textsuperscript{\textregistered} Gold 6142 at 2.6~GHz, 
with message passing techniques for parallel computing.

Figure \ref{f:dimonte} displays the power as function of the longitudinal position inside the tube.
The time average power of \textsc{dimoha}, at the pulsation $\omega$, is computed using Eq.~\eqref{e:powerHarmo} with $\hat{E_z}(z,\omega)$
being the Fourier transform in time over one period $1 / F$  of the electric field ${E_z}$ \eqref{e:VsnEsn} from the discrete model.
Figure \ref{f:dimonteelec} displays the electron velocities inside the tube at $t = 0.7$ \SI{}{\micro\second}.
The dashed line corresponds to the wave phase velocity.
The first part, up to 250 mm, is the linear growth of the wave where particles slowly lose their momentum to the benefit of the wave. Bunching of electrons starts to occur. After 250 mm, we enter progressively into the nonlinear regime where electrons, trapped in the potential well, can regain their momenta (see Appendix of Ref.~\onlinecite{min19epjh} for a pedestrian description of the phenomenon). Then, we observe five trapping periods (6 with \textsc{dimoha}) where the wave and bunched particles alternatively exchange their momenta. After 500~mm, electrons are completely trapped.

The agreement in Fig.~\ref{f:dimonte} in the linear regime (from 0 to 250 mm) and the first trapping oscillation (from 250 mm to 700 mm) is excellent, especially considering the size of the device. 
After 700 mm, discrepancies start, in particular with the trapping period being shorter with \textsc{dimoha}.
We also remark that the measured power is slightly more attenuated over the distance.
Several causes are possible. It can come from the space charge model, too unsophisticated to be accurate in strong nonlinear regime.
It can also come from an incorrect representation of the tube defects or 3D asymmetries like the beam confinement.
However, the agreement remains remarkable to allow \textsc{dimoha} to have practical applications, especially taking into account the rapidity of our algorithm.

\section{Helix geometry with the discrete model}
\label{s:sheathDIMO} 

\subsection{Section synthesis}

The discrete model, described in Section~\ref{s:dimodel}, is based on a reformulation of electromagnetic fields \eqref{e:VsnEsn}-\eqref{e:IsnHsn} that requires to express the field shape functions $\bcE^s_{n}$ and $\bcH^s_{n}$.
In this section, we extend the discrete model to 3D models.
Using a 3D helix geometry, we define the full shape functions. 
Knowledge of those functions can be used to build 3D codes that can profit from the discrete model advantages.

\subsection{The helix geometry for TWTs} 
\label{s:helix} 

The first time the helix was introduced as a metallic waveguide was in 1889 by Hertz,\cite{her89} followed by Thomson, \cite{tho93} Rayleigh,\cite{ray97} Pocklington\cite{poc97} and others. 
The sheath helix model (see below) was introduced by Ollendorff\cite{oll26} in 1926.
Since the 1940s, the helix stands as the preferred geometry\cite{min19epjh,gil11} for SWSs of TWTs, especially for communications, 
because it allows a broad bandwidth and large interaction impedance with a simple design. 
The development of helix theories and simulations is a recurrent topic in the TWT literature (for a non exhaustive list, see Refs~\onlinecite{pie47a,pie47b,cut48,chu48,pie50,sen51,
tie53,wat53,wat54,sen55,wat58,cre64,gil11,fre95,fre96,dag98,che99,pch18}) as well as for other devices like particle accelerators,\cite{chi51,chi57} free-electron lasers\cite{sha94} or gyrotrons.\cite{uhm83}
Two of the most popular models are the sheath helix model focusing on the fundamental mode of propagation, and its extension, the tape helix model, involving an infinite set of space harmonics. 
We focus on the simplest form of the sheath helix model without dielectric layers but combined with the discrete model.
Note that, as displayed in the dispersion relation
and in the interaction impedance of Fig.~\ref{f:dispRelImp}, 
the sheath helix model we use without dielectric layers differs significantly from real devices. 
Adding the missing layers improves its accuracy as shown in Ref.~\onlinecite{dim78} with four layers and Ref.~\onlinecite{des20} with five layers.
In this paper, we focus on providing a basis that can be used for 3D applications of the discrete model.

The basis for this Sub-Section originates from Refs~\onlinecite{sen55,cre64}.  
The sheath helix is combined with the discrete model in the next Sub-Section.

The helix is periodic with pitch $d$ along the $z \in \RR$ axis, extended to infinity in both directions.
The tilt is measured by the pitch angle $\Psi = \arctan \left( \frac{d}{2 \pi a} \right)$,
with $a$ the radius of the cylinder.
The system is lossless.
We consider two specific cases: 
the tape helix model, which requires taking into account all space harmonics $s$, 
and the sheath helix limited to the fundamental mode $s=0$ (we mainly focus on the latter one).
Materially, within the tape helix model, the helix is composed of thin and tilted conducting tapes, 
with $\delta$ the wire thickness, while for the sheath helix model, the helix is composed of tilted conducting sheets.

The unit vectors $\bfe_{\|}$ and $\bfe_{\bot}$, respectively parallel and perpendicular to helix wires, read
\begin{align}
\bfe_{\|} &= \bfe_z \sin \Psi + \bfe_{\theta} \cos \Psi \, ,\\
\bfe_{\bot} &= \bfe_r \times \bfe_{\|} = \bfe_z \cos \Psi - \bfe_{\theta} \sin \Psi \, ,
\end{align}
with the corresponding cylindrical coordinates $(r, \theta, z)$ and their unit vectors.
The physical electric field is spatially decomposed as $\bfE(r,\theta,z) = E_r(r,\theta,z) \, \bfe_r + E_{\theta}(r,\theta,z) \, \bfe_{\theta} + E_z(r,\theta,z) \,  \bfe_z = \sum_\ell E_\ell (r, \theta, z) \bfe_\ell (\theta)$
with $\ell \in \{r, \theta, z\}$ and similarly for the magnetic field.
Due to the axial periodicity, the electric and magnetic fields 
can be decomposed on basis fields which meet the Floquet condition \eqref{e:Floquet}.
Moreover, azimuthal and longitudinal variation of fields can be expressed\cite{sen55,cre64} as Fourier series  
\begin{align}
  E_\ell (r,\theta,z) =  \sum_{s \in \ZZ}  E_\ell^{s}(r) \upe^{-\rmi  s  \theta} \upe^{-\rmi z (\beta + 2 \pi s/ d) } \, , 
  \label{e:EfieldspaceHarmo} 
  \\
  H_\ell (r,\theta,z) = \sum_{s \in \ZZ}  H_\ell^{s}(r) \upe^{-\rmi s \theta} \upe^{-\rmi z (\beta + 2 \pi s/ d) } \, , 
  \label{e:HfieldspaceHarmo}
\end{align}
with $s \in \ZZ$ now representing the space harmonic number  
and $\beta$ the fundamental wavenumber. 
Here, the axial and angular space harmonic numbers of Refs~\onlinecite{sen55,cre64} are identical.
Amplitudes $\bfE^{s}$ and $\bfH^{s}$ are the $s$\textsuperscript{th} Fourier components of the fields.

Both tape helix and sheath helix distinguish two regions: inside the helix (superscript $\rm i$), where $r \in [0,a]$, and outside the helix (superscript $\rm o$), where $r \in [a,\infty[$. 
Assuming the wire is a perfect metallic conductor, the boundary conditions\cite{sen51} between both regions are
\begin{align}
  E^{\rm i}_{z} (a)  &= E^{\rm o}_{z} (a)  \, , 
  \label{e:BoundaryHelix1} 
  \\
   E^{\rm i}_{\theta} (a)  &= E^{\rm o}_{\theta} (a) \, ,  
   \label{e:BoundaryHelix2} 
   \\
   E^{\rm i,o}_{z} (a)  &= - E^{\rm i,o}_{\theta} (a) \cot \Psi \, , 
   \label{e:BoundaryHelix3} 
   \\
   H^{\rm i}_{z} (a) + H^{\rm i}_{\theta} (a) \cot \Psi  
   &= H^{\rm o}_{z} (a)  + H^{\rm o}_{\theta} (a) \cot \Psi \, .
   \label{e:BoundaryHelix4}
\end{align}
More complete models exist where the helix is surrounded by dielectric layers\cite{des20,dag98,che99} 
with relative permittivities $\epsilon_\rmr \neq 1$ in the $r \in [a,\infty[$ region.

\subsection{Sheath helix model for the discrete model}
\label{s:subSHelDIMO} 

In cylindrical coordinates, the electromagnetic eigenfields to be used within the discrete model are now expressed as $\bcE^s_{\beta}(r,\theta,z) =  [\cE^s_{r,\beta}(r)\bfe_{r}  +  \cE^s_{\theta,\beta}(r) \bfe_{\theta}  +   \cE^s_{z,\beta}(r)\bfe_{z} ] \upe^{-\rmi (s \theta +z (\beta + 2 \pi s/ d))} $ for each spatial component and the same for the magnetic eigenfields. 
To express each component, one solves the Helmholtz equations \eqref{e:Helmo1}-\eqref{e:Helmo2}.
We define the radial phase constant (a.k.a.\  transverse constant) as
\begin{equation}
  \Gamma = \sqrt{\beta^2 - k^2} \, ,
\end{equation}
for $s=0$, with the wavenumber in vacuum $k = \Omega_{\beta} / c$.

For the rest of this paper, we focus on the sheath helix model ($s=0$) for the sake of simplicity.\cite{wat58}

The solutions of the Helmholtz equations \eqref{e:Helmo1}-\eqref{e:Helmo2} in cylindrical coordinates are inside the helix, for $0 \leqslant r \leqslant a$, 
\begin{align}
\cE^{\rm i}_{r,\beta}(r) &=  \rmi \frac{\beta}{\Gamma} C^{\rm i}_{\beta} I_1 (\Gamma r)  \, , \label{e:erin} \\ 
\cE^{\rm i}_{\theta,\beta}(r)  &= - \rmi \frac{\Omega_{\beta} \mu_0}{\Gamma} D^{\rm i}_{\beta} I_1 (\Gamma r)  \, , \\
\cE^{\rm i}_{z,\beta}(r) &= C^{\rm i}_{\beta} I_0 (\Gamma r)  \, , \label{e:ezin} \\
\cH^{\rm i}_{r,\beta}(r) &=  \rmi \frac{\beta}{\Gamma} D^{\rm i}_{\beta} I_1 (\Gamma r) \, , \\ 
\cH^{\rm i}_{\theta,\beta}(r) &=   \rmi \frac{\Omega_{\beta} \epsilon_0}{\Gamma}  C^{\rm i}_{\beta} I_1 (\Gamma r) \, , \\
\cH^{\rm i}_{z,\beta}(r) &= D^{\rm i}_{\beta} I_0 (\Gamma r) \, ,
\end{align}
and outside the helix, for $a \leqslant r < \infty$, 
\begin{align}
\cE^{\rm o}_{r,\beta}(r) &= - \rmi \frac{\beta}{\Gamma} C^{\rm o}_{\beta} K_1 (\Gamma r) \, , \label{e:erout} \\
\cE^{\rm o}_{\theta,\beta}(r) &=  \rmi \frac{\Omega_{\beta} \mu_0}{\Gamma} D^{\rm o}_{\beta} K_1 (\Gamma r)  \, , \\
\cE^{\rm o}_{z,\beta}(r) &= C^{\rm o}_{\beta} K_0 (\Gamma r) \, , \label{e:ezout}  \\
\cH^{\rm o}_{r,\beta}(r) &= - \rmi \frac{\beta}{\Gamma} D^{\rm o}_{\beta} K_1 (\Gamma r) \, , \\
\cH^{\rm o}_{\theta,\beta}(r) &= - \rmi \frac{\Omega_{\beta} \epsilon_{0}}{\Gamma}  C^{\rm o}_{\beta} K_1 (\Gamma r) \, , \\
\cH^{\rm o}_{z,\beta}(r) &= D^{\rm o}_{\beta} K_0 (\Gamma r)  \, ,  \label{e:hzout}
\end{align}
with arbitrary constants $C^{\rm i}_{\beta}$, $C^{\rm o}_{\beta}$, $D^{\rm i}_{\beta}$, $D^{\rm o}_{\beta}$ computed below.
Factors $I_s$ and $K_s$ are modified Bessel functions\cite{abr64} of $s$\textsuperscript{th} order of the first and the second kinds respectively.
We recall the useful relations\cite{abr64} $I_{0} (0)=1$, $I_{s\neq 0} (0)=0$, 
$\partial_x I_0(x) = I_1(x)$, 
and $\partial_x K_0(x) = - K_1(x)$. 
Note that $I_0 (0) =1$ implies $\cE^{\rm i}_{z,\beta}(r = 0)  = C^{\rm i}_{\beta}$ in Eq.~\eqref{e:ezin}.

The link between the constants is given by the boundary equations \eqref{e:BoundaryHelix1}-\eqref{e:BoundaryHelix4}, applied on Eqs~\eqref{e:erin}-\eqref{e:hzout} evaluated at $r = a$.
This yields
\begin{align}
C^{\rm o}_{\beta} &= \frac{I_0 (\Gamma a)}{K_0(\Gamma a)} \, C^{\rm i}_{\beta} \, , \label{e:bondaBessel1} \\
D^{\rm o}_{\beta} &= - \frac{I_1 (\Gamma a)}{K_1(\Gamma a)} \, D^{\rm i}_{\beta} \, , \label{e:bondaBessel2} \\
D^{\rm i}_{\beta} &= - \frac{I_0(\Gamma a)}{I_1(\Gamma a)} \frac{\rmi \Gamma}{\Omega_{\beta} \mu_0} \tan (\Psi) \, C^{\rm i}_{\beta} \, , \label{e:bondaBessel3} \\
D^{\rm o}_{\beta} &=  \frac{K_0(\Gamma a)}{K_1(\Gamma a)} \frac{\rmi \Gamma}{\Omega_{\beta} \mu_0} \tan (\Psi) \, C^{\rm o}_{\beta} \, , \label{e:bondaBessel4} 
\end{align}
and the determinantal equation 
\begin{align}
\left( \tan \Psi \right)^2 &= \frac{I_1(\Gamma a) K_1(\Gamma a)}{I_0(\Gamma a) K_0(\Gamma a)} \frac{ (\Omega_{\beta} )^2 \epsilon_0 \mu_0}{\Gamma^2} \, . \label{e:bondaBessel5}
\end{align}
Given the radius $a$ and the dispersion relation \eqref{e:bondaBessel5} of the helix, we only need to find the values of $C^{\rm i}_{\beta}=\cE^{\rm i}_{z,\beta}(0)$ 
to express completely eigenfields Eqs~\eqref{e:erin}-\eqref{e:hzout}.

A major advantage of the sheath helix model is that it only requires the value of the pitch $d$ and the helix radius $a$ to fully describe the structure.
However, as displayed in Fig.~\ref{f:dispRelImp}, this model is too incomplete to accurately reproduce experimental helix TWTs. Real devices are composed of concentric layers of dielectric tapes, glass, vacuum or air with different permittivities.
Those layers add new boundary conditions modifying the eigenfields in the $r$ and $\theta$ directions. 
In this paper, our aim is not to provide the most accurate representation of the helix since it can change according to devices, but rather to provide a canvas for later investigations of more realistic geometries.

\begin{figure}
  \centering
  \includegraphics[width=\columnwidth]{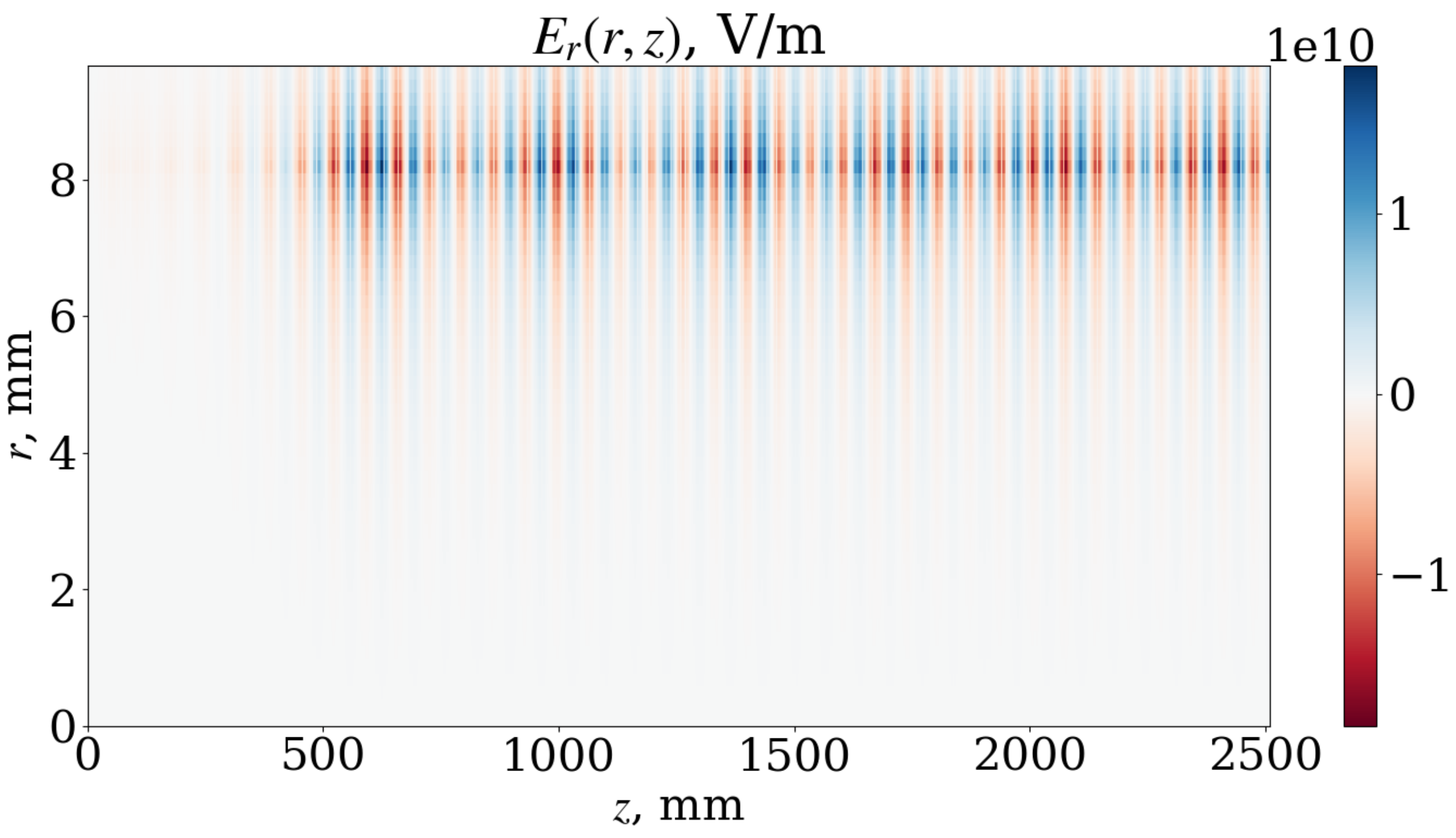} \\
  \includegraphics[width=\columnwidth]{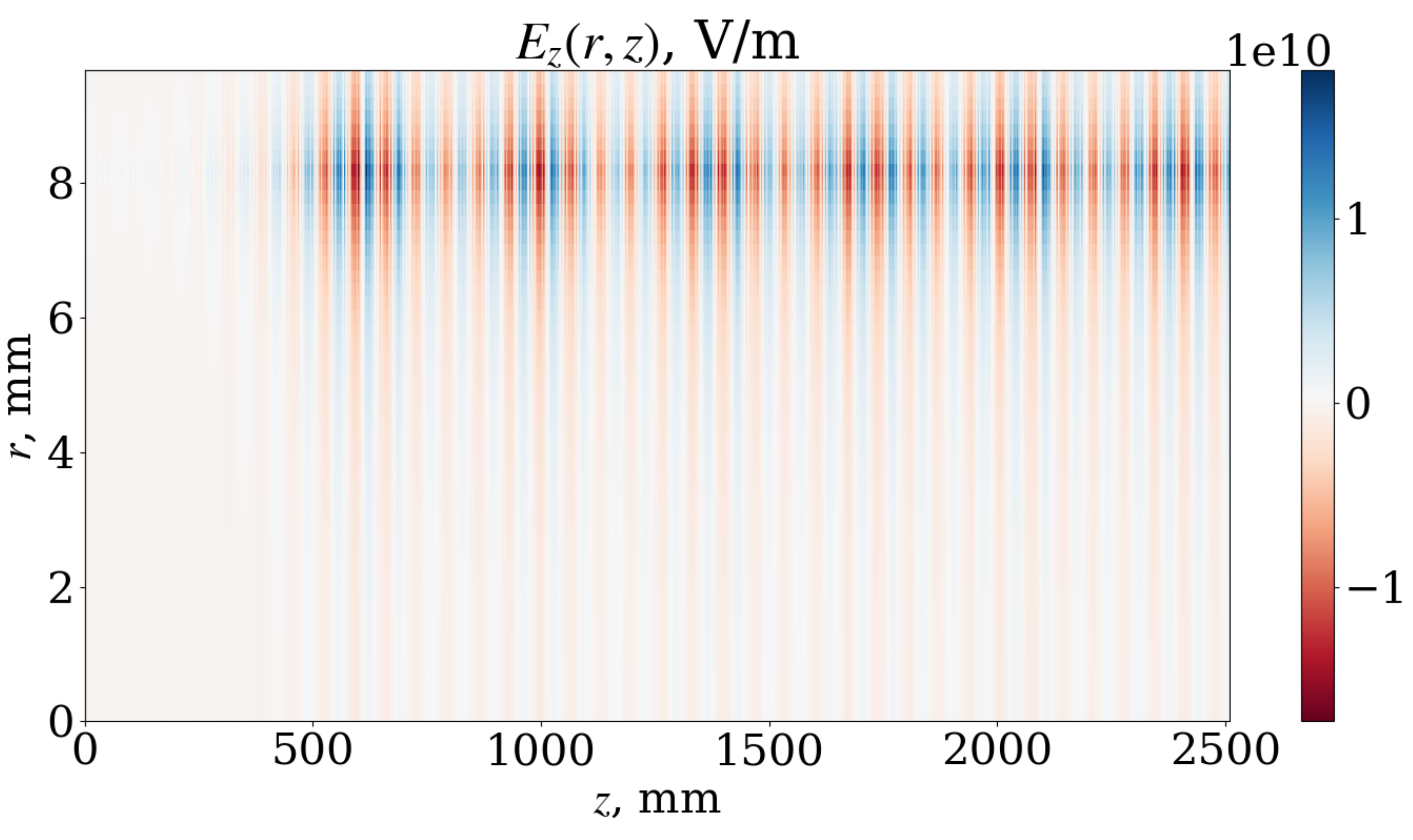} \\
  \includegraphics[width=\columnwidth]{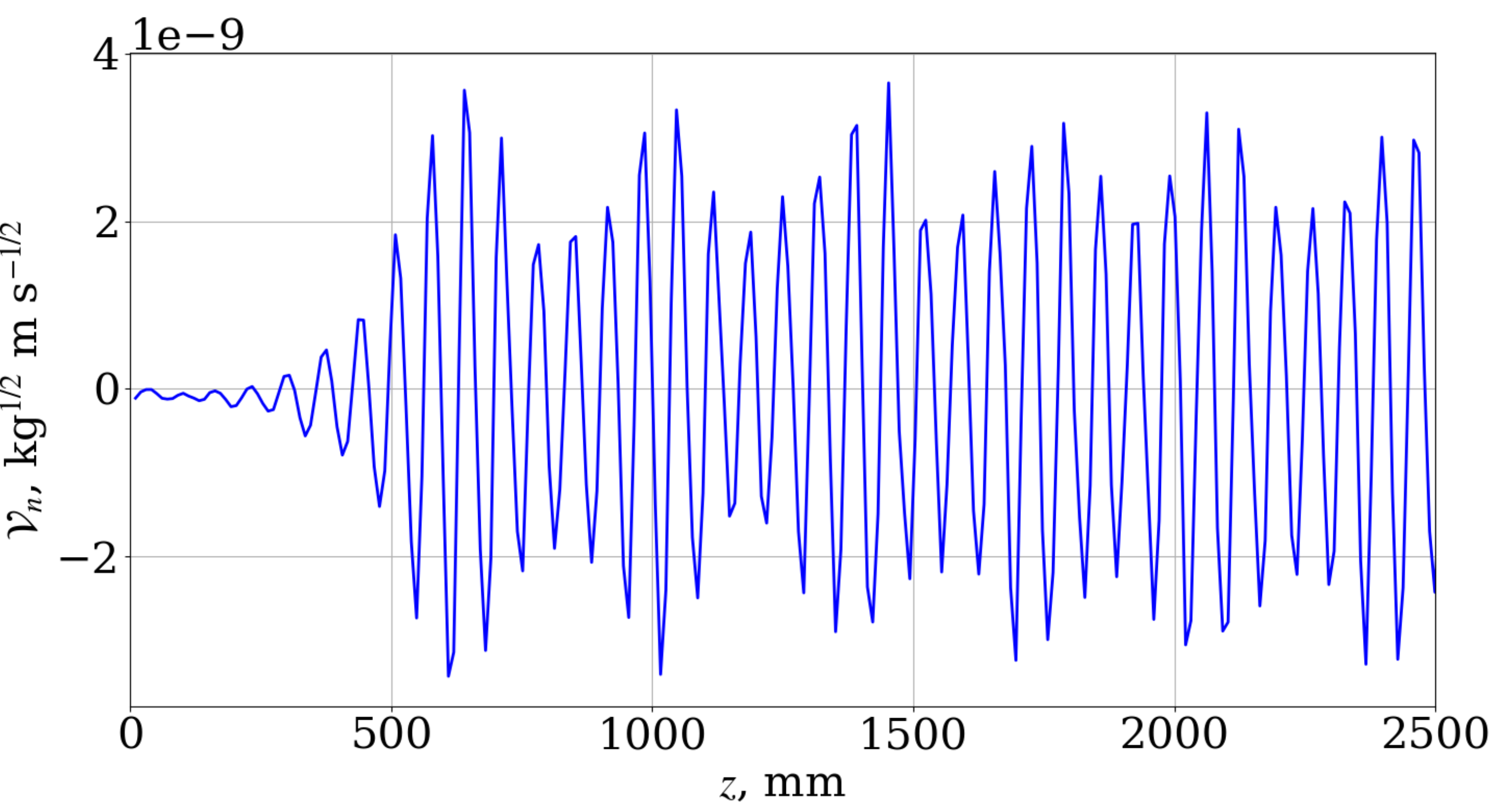} \\
  \caption{Reconstruction of the components 
    $E_r$ (first panel) and $E_z$ (second panel) of the three-dimensional real electric field 
    inside a 3 m long TWT (with helix radius $a = 8.06$~mm) at time $t = 0.7$~\SI{}{\micro\second}. 
    The electric field is given by Eq.~\eqref{e:VsnEsn}. 
    The time amplitudes $\cV^s_n$ (third panel) 
    are obtained from the 1D wave-particle interaction with \textsc{dimoha} 
    (same simulation as for Figs \ref{f:dimonte} and \ref{f:dimonteelec}), 
    and the line is a guide for the eye (the effective cell has length 10.16~mm). 
    The shape functions $\bcE^s_n$ are computed from the Gel'fand inverse $\beta$-transform 
    of the electric eigenfields given by Eqs~\eqref{e:erin}-\eqref{e:ezin}, \eqref{e:erout}-\eqref{e:ezout} 
    from the sheath helix model. 
    Those eigenfields are computed using the sheath helix dispersion relation and impedance.} 
  \label{f:EMfields}
\end{figure}

\subsection{Longitudinal eigenfield constants}
\label{s:EsbFrompower}

Now, we want to express the coefficients $C^{\rm i}_{\beta}$.
Since eigenfields $\bcE^{s}_{\beta}(\bfr),\bcH^{s}_{\beta}(\bfr)$ are time-independent 
(and so are the $C^{\rm i}_{\beta}$'s), 
we can work in harmonic regime where fields are decomposed as 
$\bfE(\bfr,t) = \Re [ \tilde{\bfE}(\bfr) \upe^{\rmi \omega t} ]$.
In this regime, the average flux of the Poynting vector\cite{jac99} (average electromagnetic power) is
$ \langle P_z \rangle 
  = 1/2 \int_{\cS}  \Re \left( \tilde{\bfE}^* \times \tilde{\bfH} \right)   \cdot \bfe_z  \upd \cS$,
where $\cS$ is the waveguide transverse section ($|\cS| = \pi a^2$ for the inner part of the helix).
We rewrite this power in the $\beta$-based representation using the field decomposition Eqs~\eqref{e:VbetaEbeta}-\eqref{e:IbetaHbeta} as
\begin{equation} 
  \langle P_{z} \rangle  
  = \frac{1}{2} \sum_{s \in \ZZ} \Re \frac{1}{2 \pi}  \int^{\pi}_{-\pi} 
           \tilde{\cV}^{s*}_{\beta} \rmi \tilde{\cI}^{s}_{\beta} \frac{1}{d} \sK^s_{\beta} c^2   \upd (\beta d)  \, .
  \label{e:HarmoPowerBeta}
\end{equation}
From Maxwell equations without sources,
and thanks to the eigenfields orthogonality, 
we have\cite{min19ps} $\tilde{\cV}^{s}_{\beta} = \rmi \tilde{\cI}^{s}_{\beta}$.
We define the refractive kernel\cite{min18}
\begin{equation}
  \sK^s_{\beta} =  \frac{v_{\rmg} \, \Omega^s_{\beta}}{c^2}   = \frac{d}{c^2} \int_{\cS} \Re \big( \bcE^{s*}_{\beta} \times \bcH^{s}_{\beta} \big) \cdot \bfe_z \, \rmd^2 \bfr  \label{e:Ksbeta2}  \, ,
\end{equation} 
where the group velocity
\begin{align} 
v_{\rmg} (\beta) & = \frac{d}{\Omega^s_{\beta}} \int_{\cS} \Re \big( \bcE^s_{\beta} \times \bcH^{s*}_{\beta} \big) \cdot \bfe_z \, \rmd^2 \bfr  \label{e:vgS} \, ,
\end{align}
is expressed in terms of the flux of the Poynting vector through the section of the waveguide, 
divided with the mode energy \eqref{e:normal} in this cell.  Equation \eqref{e:vgS} is obtained from the derivative of Eq.~\eqref{e:normal}. 

Considering only the fundamental harmonic $s=0$, 
the cross product in Eq.~\eqref{e:Ksbeta2} reduces to 
\begin{align}
\frac{\sK_{\beta} \, c^2}{d} & = 2 \pi \int^{a}_{0} \Re \left( \cE^{\rm i*}_{r,\beta} \cH^{\rm i}_{\theta,\beta} - \cE^{\rm i*}_{\theta,\beta} \cH^{\rm i}_{r,\beta} \right) r \upd r  \nonumber \\
 & \quad \,  + 2 \pi \int^{\infty}_{a} \Re \left( \cE^{\rm o*}_{r,\beta} \cH^{\rm o}_{\theta,\beta} - \cE^{\rm o*}_{\theta,\beta} \cH^{\rm o}_{r,\beta}  \right) r \upd r \, , \label{e:Kbcodint}
\end{align}
where the integral over $\theta$ is equal to $2\pi$. 
With the eigenfield solutions \eqref{e:erin}-\eqref{e:hzout}, 
and the boundary conditions Eqs~\eqref{e:bondaBessel1}-\eqref{e:bondaBessel5}, 
Eq.~\eqref{e:Kbcodint} reduces to
\begin{equation}
   \frac{\sK_{\beta} c^2}{d}  
   = \frac{\beta \Omega_{\beta} \epsilon_0}{\Gamma^2} \pi a^2 \, |C^{\rm i}_{\beta}|^2 \, F(\Gamma a) \, ,   
   \label{e:Kbeta0Integral}
\end{equation}
with the dimensionless impedance reduction factor\cite{cut48,pie50}
\begin{align}
  F(\Gamma a) 
  & =  \left( 1 + \frac{I_0 K_1 }{I_1 K_0 } \right) \left( I^2_1 - I_0  I_2  \right) \nonumber \\ 
  &\quad \, + \left( \frac{I_0}{K_0} \right)^2 \left( 1 + \frac{I_1 K_0}{I_0 K_1 } \right) \left( K_0 K_2 - K^2_1  \right) \, ,
\end{align}
with all Bessel functions evaluated at $\Gamma a$.

Before continuing, we must express the helix impedance from the power.
For vacuum electron tubes, the electromagnetic power is generally computed for one frequency in harmonic regime (this is the time average power) from the 1D telegrapher's equations\cite{pie50} as
\begin{equation}
  \langle P_{z} \rangle (z,\omega) 
  = \frac{|\hat{E_z}(z,\omega)|^2}{2 \beta^2 Z_{\mathrm{c}}(\beta)}  \, , 
  \label{e:powerHarmo}
\end{equation}
in the monochromatic (a.k.a.\ continuous waveform, CW) regime, and must be equal to Eq.~\eqref{e:HarmoPowerBeta}.
The peak electric field $\hat{E_z}$ is linked to the longitudinal field $E_z (z) = \Re [ \hat{E_z} \rme^{- \rmi \beta z}]$, where $\Re$ denotes the real part.

The projection on the $z$ axis of the electric field amplitude from Eq.~\eqref{e:VbetaEbeta} yields $ |\tilde{E}_{z}(0)|^2  = 1/(2 \pi) \int^{\pi}_{-\pi} \tilde{\cV}_{\beta} \tilde{\cV}^{*}_{\beta}  | C^{\rm i}_{\beta} |^2  \upd (\beta d)$, 
in the monochromatic regime. 
Substituting it into the monochromatic power \eqref{e:powerHarmo} and \eqref{e:HarmoPowerBeta}, and knowing Eq.~\eqref{e:Kbeta0Integral}, 
leads to the coupling (interaction) impedance
\begin{equation} 
   Z_{\mathrm{c}}(\beta) 
   = \frac{1}{ \pi  a^2 \epsilon_0} \frac{ \Gamma^2}{\Omega_{\beta} \beta^3} \frac{1}{F(\Gamma a)} 
   \label{e:3DZc}
\end{equation}
established by Pierce.\cite{pie50}  
Similarly to the dispersion relation, a more accurate calculation of the coupling impedance 
is obtained using more detailed models of actual helices, 
as shown in Refs~\onlinecite{dim78,des20}. 
 
The electric eigenfunction on the $z$-axis is obtained from Eqs~\eqref{e:3DZc} and \eqref{e:Kbeta0Integral} (with $c^2 \sK_{\beta} = \Omega_{\beta} v_{\rmg}$) as
\begin{equation}
   \cE^{\rm i}_{z,\beta}(0) =  C^{\rm i}_{\beta} 
   = \sqrt{\beta^2  \frac{v_{\rmg} \Omega_{\beta}}{d} Z_{\mathrm{c}}(\beta)  } \, .
   \label{e:EsbFromHelix}
\end{equation}
This 1D electric eigenfield \eqref{e:EsbFromHelix} was already obtained\cite{ber11,min18} 
(see Eq.~\eqref{e:Ebeta1D}) from a 1D approximation of the electromagnetic power and should be valid for any periodic geometry.
Since $C^{\rm i}_{\beta}$ is time-independent, this relation is true in both the frequency and time domains.  

Now, we have all the terms to express the electric and magnetic eigenfields \eqref{e:erin}-\eqref{e:hzout} within the sheath helix model.
Using the Gel'fand inverse $\beta$-transform \eqref{e:betainverse}, we can express all of the shape functions $\cE^s_{r,n} (r,\theta,z),\cE^s_{\theta,n}(r,\theta,z),\cE^s_{z,n}(r,\theta,z)$ for the electric field, $\cH^s_{r,n}(r,\theta,z),\cH^s_{\theta,n} (r,\theta,z),\cH^s_{z,n}(r,\theta,z)$ for the magnetic field, and $\cA^s_{r,n} (r,\theta,z),\cA^s_{\theta,n} (r,\theta,z),\cA^s_{z,n}(r,\theta,z)$ for the vector potential.
To find the real fields \eqref{e:VsnEsn}, \eqref{e:IsnHsn} and \eqref{e:potVec}, 
we only need to obtain the time-dependent amplitudes $\cV^s_n$ and $\cI^s_n$, 
which are computed from our 1D multi-particle dynamics \eqref{e:evo1}-\eqref{e:evo3} with \textsc{dimoha}.

Figure \ref{f:EMfields} shows a reconstruction of full electric field \eqref{e:VsnEsn} in the 3~m long TWT. The magnetic field can also be reconstructed.
Note that the modified Bessel functions of the first kind $I_s$ (used for eigenfields inside the helix) are increasing functions 
while the modified Bessel functions of the second kind $K_s$ (used for eigenfields outside the helix) are decreasing functions: 
the RF fields are concentrated near the conducting helix.
We use the time amplitudes $\cV^s_n$ obtained from the 1D wave-particle interaction 
with \textsc{dimoha} (same simulation as for Figs \ref{f:dimonte} and \ref{f:dimonteelec}).
From it, thanks to the decomposition of the discrete model, we can reconstruct the fields ; 
for this proof-of-principle, we do not compute eigenfunctions with a detailed several-media model  but we use the sheath helix fields for simplicity, as Fig.~\ref{f:dispRelImp} shows a rough general agreement between actual characteristics of the device and this elementary model. 
The same approach can be used to reconstruct the magnetic field and the vector potential.
The main interest with this is that we now have a 3D representation of fields allowing us to build a 3D wave-particle dynamics. 
In 3D, the number of DOFs of the fields ($\cV^s_n$, $\cI^s_n$) remains the same as in 1D. Therefore, the discrete model is a promising theory for such 3D codes.
Note that the construction of 3D codes requires a 3D propagation model for particles that includes the space charge.

\section{Electromagnetic power with the discrete model}
\label{s:power} 

\subsection{Section synthesis}

In this section, we express the electromagnetic power using the discrete model. 
Two expressions are proposed: a monochromatic approximation and an explicit non-monochromatic expression.

\subsection{Total electromagnetic power}

The electromagnetic power is a prominent feature of vacuum electron tubes. 
However, this power is generally computed in harmonic regime from a 1D model 
using Eq.~\eqref{e:powerHarmo} in the monochromatic regime. 
As an alternative, the second main purpose of this paper is to compute the power in the time domain using the discrete model.
Indeed, time-domain expressions are well suited for observing, for example, 
wave oscillations and transient reflections on the power, 
which would be impossible with frequency models.

In frequency domain, we already expressed the power with the discrete model (see Eq.~\eqref{e:HarmoPowerBeta}).
With the decomposition Eqs~\eqref{e:VsnEsn}-\eqref{e:IsnHsn} of the discrete model, the electromagnetic power becomes
\begin{equation}
  P_z(z,t) 
  = \sum_{s \in \ZZ} \sum_{n_1, n_2} \, \cV^{s}_{n_1}(t) \, \cI^{s}_{n_2}(t) \, \frK^{s}_{n_1, n_2}(z) \, ,
  \label{e:PVIK}
\end{equation}
with
\begin{equation} \label{e:ksn1n2}
\frK^{s}_{n_1, n_2}(z)
   =   \int_\cS \left( \bcE^{s}_{-n_1}(\bfr) \times \rmi \bcH^{s}_{-n_2}(\bfr) \right) \cdot \bfe_z \, \rmd \cS \, .
\end{equation}
To compute $\frK^{s}_{n_1, n_2}$, 
we need the $\bfr$-dependence of the basis fields $\bcE^{s}_{n_1}, \bcH^{s}_{n_2}$,
or equivalently of eigenfields $\bcE^{s}_{\beta_1}, \bcH^{s}_{\beta_2}$. 
For a helical waveguide, the field dependence reduces to a phase factor along with azimuthal rotation, 
\begin{equation}
  \bcE^s_\beta (r, \theta, z) 
  = \rme^{- \rmi (z (\beta + 2 \pi s/ d)+ s \theta)} \, \cR_{\theta + 2 \pi z/d} \, \bcE^s_\beta (r, 0, 0) \, ,
  \label{e:helixE}
\end{equation}
where $\cR_\alpha$ is the rotation operator with angle $\alpha$ about the $z$-axis,
and similarly for $\bcH^s_\beta$.
In the $\beta$-based representation, Eq.~\eqref{e:ksn1n2} then reads
\begin{multline}
  \frK^{s}_{n_1, n_2} (z) 
  = \frac{1}{(2 \pi)^{2}} \,  \int_{- \pi}^\pi  \int_{- \pi}^\pi  \rmi \Big( \frk^{s}_{\beta_1, \beta_2} \\
      \cdot \rme^{\rmi \beta_1 (z - n_1 d) - \rmi \beta_2 (z - n_2 d)} \Big) \,  \rmd (\beta_1 d) \, \rmd (\beta_2 d) \, , \label{e:kn1n2kb1b2}
\end{multline}
with 
\begin{equation}
\frk^s_{\beta_1, \beta_2}
  = \int_\cS \Big( \bcE^{s*}_{\beta_1} (r, \theta, 0) \times \bcH^{s}_{\beta_2}(r, \theta, 0) \Big) 
               \cdot \bfe_z \, \rmd \cS \, ,
  \label{e:kb1b2}
\end{equation}
with $\rmd \cS = r \rmd \theta \rmd r$ in cylindrical coordinates. Rotation operators from \eqref{e:helixE} disappear in \eqref{e:kb1b2} thanks to the integration over the disk $\cS$. 
The reciprocity relations imply that $\frk^s_{- \beta_1, - \beta_2} = - \frk^{s *}_{\beta_1, \beta_2}$.
Moreover, for $\beta_2 = \beta_1$, the integral Eq.~\eqref{e:kb1b2} relates directly 
with the group velocity Eq.~\eqref{e:vgS} and the refractive kernel Eq.~\eqref{e:Ksbeta2} 
as $\frk^s_{\beta, \beta} =  v_{\rmg} \Omega^s_{\beta} / d =  c^{2} \sK^s_{\beta} /d$.

\subsection{Electromagnetic power per cell}
 
\begin{figure}
\centering
\includegraphics[width=\columnwidth]{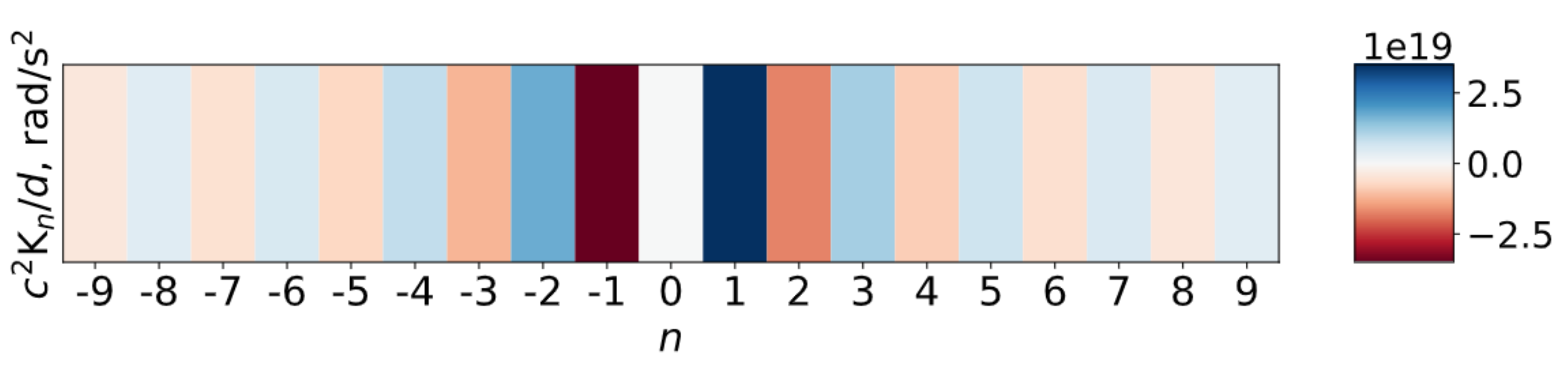}
\caption{Monochromatic refractive kernel $\sK_{n}$ \eqref{e:ksntransform}.
We use the dispersion relation from the sheath helix model displayed in Fig.~\ref{f:dispRelImp} 
with pitch $d = 2.54$~mm and radius $a = 8.06$~mm.}
\label{f:Kn}
\end{figure}

\begin{figure}
\centering
\includegraphics[width=\columnwidth]{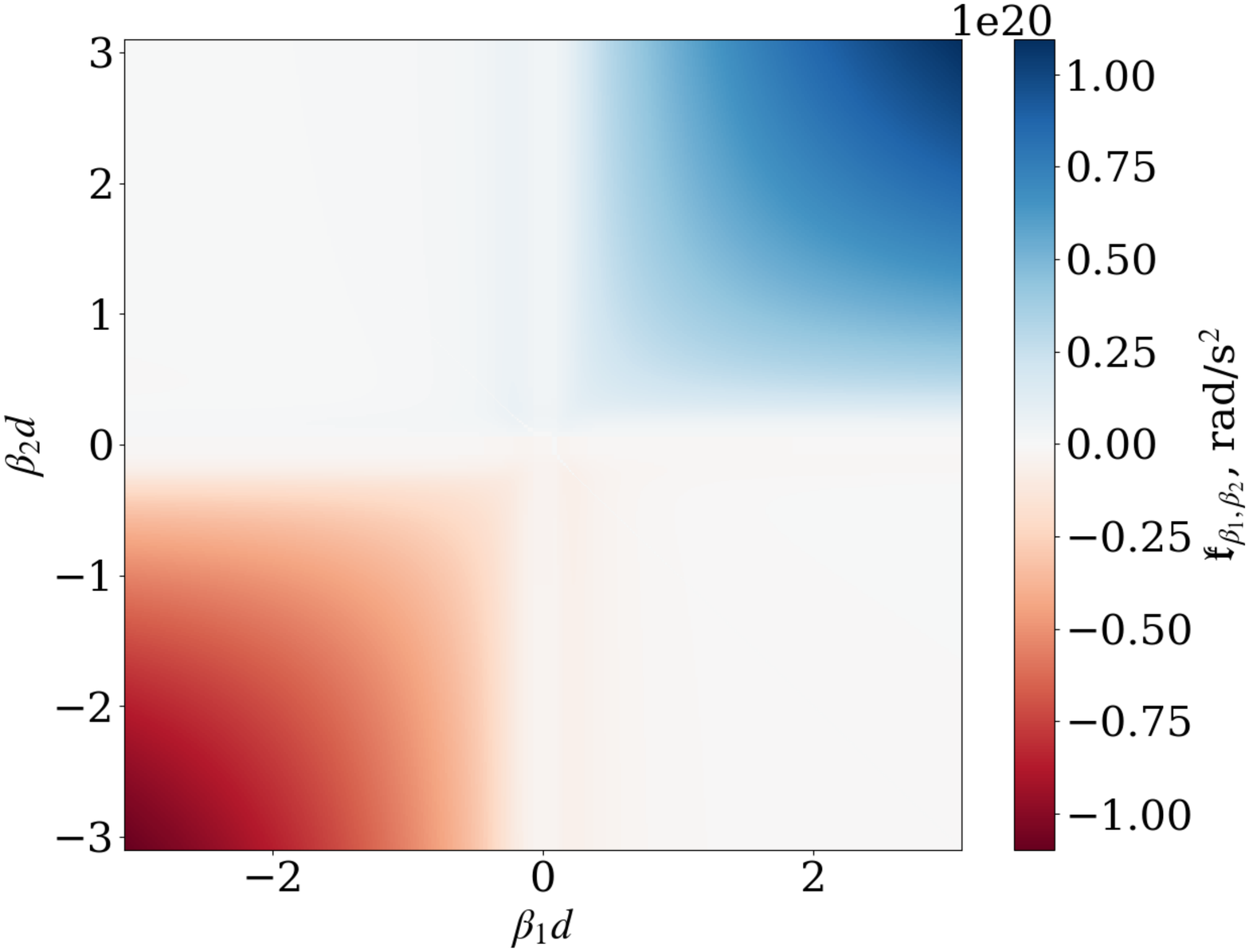} 
\caption{Coefficient $\frk_{\beta_1, \beta_2}$, from Eq.~\eqref{e:finalfrk}, 
as function of phases $\beta_1 d$ and $\beta_2 d$ from the sheath helix model 
with pitch $d = 2.54$~mm and radius $a = 8.06$~mm. 
We verified that, when $\beta_1 = \beta_2$, we have $\frk^s_{\beta_1, \beta_1} =  v_{\rmg} \Omega^s_{\beta_1} / d =  c^{2} \sK^s_{\beta_1} /d$.}
\label{f:Kbb}
\end{figure}

\begin{figure}
\centering
\includegraphics[width=\columnwidth]{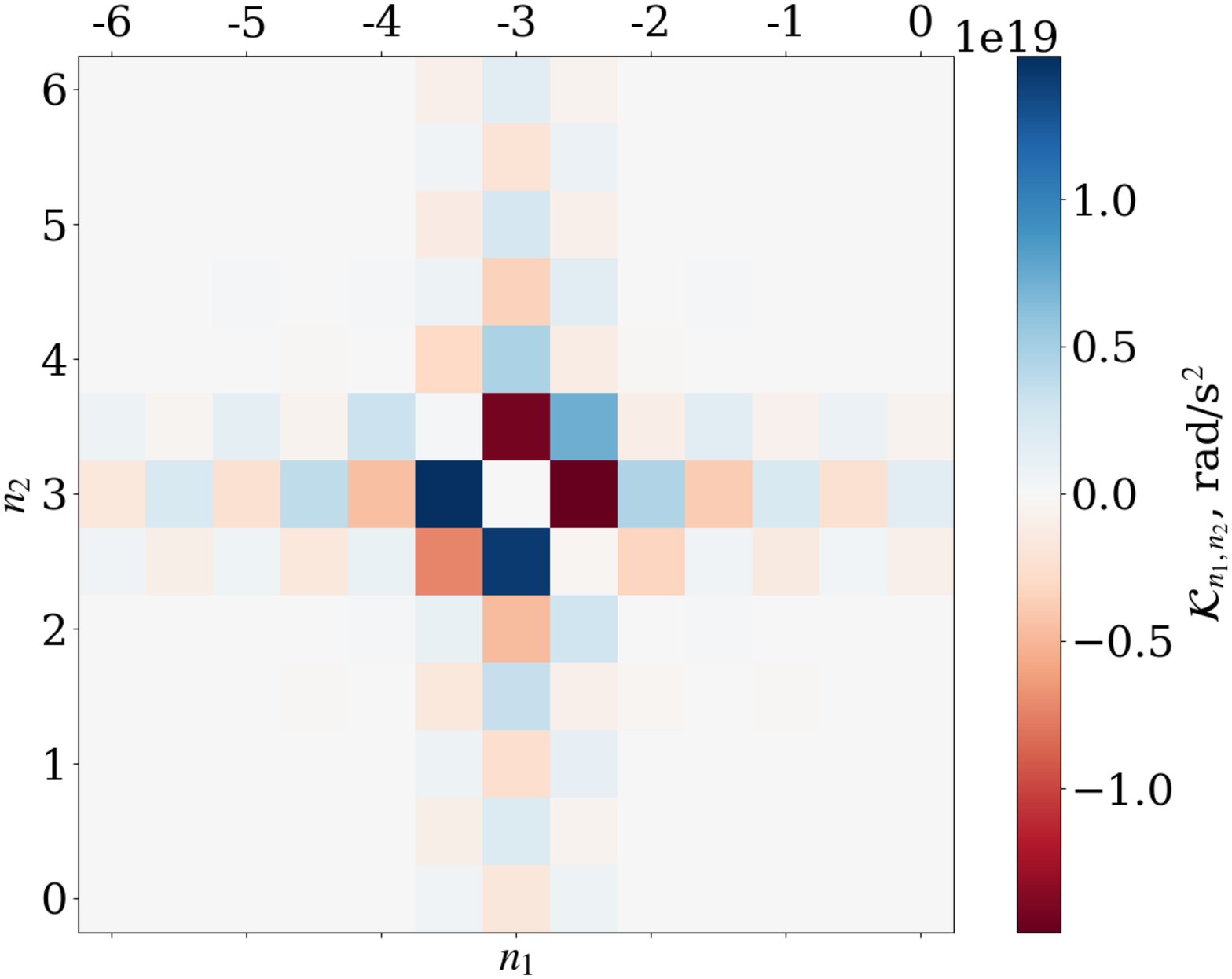}
\caption{Coefficients $\cK_{n_1, n_2}$ \eqref{e:finalKn1n2chroma} 
computed from coefficients $\frk_{\beta_1, \beta_2}$ given by \eqref{e:finalfrk} 
within the sheath helix model with pitch $d = 2.54$~mm and radius $a = 8.06$~mm.}
\label{f:Knn}
\end{figure}

Since the discrete model provides a useful representation per cell of quantities, 
we define the contribution of period $n$ to the electromagnetic power as
\begin{equation}
  P_{z, n} (t) 
  = \sum_{s \in \ZZ} \sum_{n_1, n_2} \cV^{s}_{n_1}(t) \, \cI^{s}_{n_2}(t) \,     \cK^{s}_{n_1 - n, n_2 - n} \, ,
  \label{e:PnK}
\end{equation}
the average flux of Eq.~\eqref{e:PVIK}, with the geometric factor  
\begin{equation}
  \cK^{s}_{n_1 - n, n_2 - n} 
  = \frac{1}{d} \, \int_{(n-1/2)d}^{(n+1/2)d} \frK^{s}_{n_1, n_2} (z) \, \rmd z \, .
\label{e:sK}
\end{equation}

Note that the integral in Eq.~\eqref{e:sK} can be seen as the integral of the rectangular (or gate) function $\Pi$ 
over the total length of the structure as 
$\int_{(n-1/2)d}^{(n+1/2)d} \cG (z) \, \rmd z = \int_{-\infty}^{\infty} \Pi (z/d -n) \cG (z) \, \rmd z$ 
for any $\cG$, with $\Pi(u) = 1$ if $|u| < 1/2$, and 0 otherwise.
Since the Fourier transform of the rectangular function is 
\begin{multline}
  \int_{-\infty}^{\infty} \Pi (z/d -n) \, \rme^{-\rmi ( \beta_2 - \beta_1 ) z} \, \rmd z 
  \\
  =  \sinc_\pi \left( \frac{d \, ( \beta_2 - \beta_1 )}{2 \pi}\right) \, \rme^{-\rmi n d( \beta_2 - \beta_1 )}  \, ,
\end{multline}
with the normalised cardinal sine function $\sinc_\pi (x) = (\sin \pi x) / (\pi x)$, 
we can re-express Eq.~\eqref{e:PnK} in the $\beta$-based representation as
 \begin{multline}
  P_{z, n} (t) 
  = \sum_{s \in \ZZ} \frac{1}{(2 \pi)^2} 
       \int_{- \pi}^\pi \int_{- \pi}^\pi  \cV^{s*}_{\beta_1}(t) \, \cI^{s}_{\beta_2}(t) \, \rmi \, \frk^s_{\beta_1, \beta_2} 
       \\
      \cdot   \sinc_\pi \left( \frac{d ( \beta_2 - \beta_1 )}{2 \pi} \right) \, \rme^{-\rmi n d( \beta_2 - \beta_1 )} \,  
          \rmd (\beta_1 d) \, \rmd (\beta_2 d)  \, .
  \label{e:PnKinbeta}
\end{multline}
Equation~\eqref{e:PnKinbeta} involves two wavenumbers $\beta_1$ and $\beta_2$ 
from the Fourier transforms on $n_1$ and $n_2$.
We cannot impose $\beta_1 = \beta_2$ by invoking destructive interference between different wavenumbers, 
because we consider a single cell $(n-1/2) d < z < (n+1/2) d$, not the full line $-\infty < z < \infty$. 
Indeed, if we directly force $\beta_1 = \beta_2$, we lose variable $n$ in Eq.~\eqref{e:PnKinbeta} 
and our so-called power per cell reduces to two sums (on $n_1$ and $n_2$) 
over the total number of wave\-guide periods regardless of the chosen cell, 
which makes no sense. 
Therefore, to express the electromagnetic power per cell, we need to express the geometric factor
\begin{align}
   \cK^{s}_{n_1, n_2} 
   &= \frac{1}{(2 \pi)^{2}} \, \int_{- \pi}^\pi  \int_{- \pi}^\pi  \rmi \frk^{s}_{\beta_1, \beta_2 }  \sinc_\pi \left( \frac{d ( \beta_2 - \beta_1 )}{2 \pi} \right) 
   \nonumber \\
&\quad \, \cdot  \rme^{-\rmi n_1 d\beta_1}  \rme^{\rmi n_2 d\beta_2}  \,  \rmd (\beta_1 d) \, \rmd (\beta_2 d) \, . \label{e:finalKn1n2chroma}
\end{align}
Two proposals for $\cK^{s}_{n_1, n_2}$ are provided: 
a monochromatic approximation and an explicit non-monochromatic expression.

\subsection{Monochromatic approximation of the power}

We propose  a 1D approximation to Eq.~\eqref{e:PnK} in monochromatic regime (that has a lot of applications), 
where the field is a sinusoidal wave with a single pulsation $\omega = 2 \pi F$.
This expression requires only the dispersion relation.

From Eq.~\eqref{e:finalKn1n2chroma}, with a change of variables, we have
\begin{align}
  &\cK^{s}_{n_1, n_2}  = \nonumber \\
  &
  \frac{1}{(2 \pi)^2} \frac{1}{2} \Bigg[
      \int_{-\pi}^\pi \int_{\beta_1 d - \pi}^{\beta_1 d + \pi}  \rmi  \frk^{s}_{\beta_1, \beta_1 + \Delta \beta}   
          \,  \sinc_\pi  \left( \frac{\Delta \beta d}{2\pi} \right) 
      \nonumber \\
  &\qquad  \qquad \cdot 
     \, {\rm e}^{\rmi (n_2 - n_1) \beta_1 d}  \, \rme^{\rmi n_2 d \Delta \beta}  
         {\rm d} (\Delta \beta d) \, {\rm d} (\beta_1 d)  \nonumber \\
  &\qquad
  + \int_{-\pi}^\pi \int_{\beta_2 d - \pi}^{\beta_2 d + \pi}  \rmi  \frk^{s}_{\beta_2 - \Delta \beta, \beta_2}  
           \,  \sinc_\pi  \left( \frac{\Delta \beta d}{2\pi} \right) 
 \nonumber \\
 &\qquad \qquad \cdot  \, {\rm e}^{\rmi (n_2 - n_1) \beta_2 d}  \, \rme^{\rmi n_1 d \Delta \beta}
{\rm d} (\Delta \beta d) \,  {\rm d} (\beta_2 d)  \Bigg]  \, .
\label{e:K_appendix}
\end{align}
To estimate it for monochromatic signals, where $\beta_1 \simeq \beta_2$, 
we note that the integral over $\Delta \beta d$ has a limited range. 
We approximate Eq.~\eqref{e:K_appendix} with
\begin{equation}
  \sinc_\pi   \left( \frac{\Delta \beta d}{2\pi} \right) 
  \approx \sinc_\pi (0)  = 1  \, 
\end{equation} 
and we replace $\frk^{s}_{\beta_1, \beta_1 + \Delta \beta}$ 
with $\frk^{s}_{\beta_1, \beta_1} = c^{2} \sK^s_{\beta_1} /d $
using the refractive kernel \eqref{e:Ksbeta2}.
We define the modified (with the $\rmi$ factor) Fourier transform of the refractive kernel \eqref{e:Ksbeta2} as\cite{min18}
\begin{equation}
  \sK^s_{n} 
      = - \sK^s_{-n}  = \frac{1}{2\pi} \int^{\pi}_{-\pi} \rmi \sK^s_{\beta} \, \rme^{- \rmi n \beta d} \, \rmd(\beta d) \, . \label{e:ksntransform}
\end{equation}
Since
$\int_{\phi - \pi}^{\phi + \pi} \exp (- \rmi n_2 d \Delta \beta) \, \rmd (d \Delta \beta) 
= 2 \pi \delta_{n_2}^0$, 
we obtain
\begin{align}
  \cK^{s}_{n_1, n_2} 
\approx & \int^\pi_{-\pi} \frac{c^2}{d}  \rmi \sK^s_{\beta_1}  \frac{ \delta_{n_2}^0 }{4 \pi} 
  {\rm e}^{\rmi (n_2 - n_1) \beta_1 d} \, {\rm d} (\beta_1 d)  
 \nonumber \\
&
  + \int^\pi_{-\pi}  \frac{c^2}{d} \rmi \sK^s_{\beta_2}  \frac{  \delta_{n_1}^0 }{4 \pi} \, {\rm e}^{\rmi (n_2 - n_1) \beta_2 d} \, {\rm d} (\beta_2 d)      \\
\approx & \    \frac{c^2}{d} \sK^s_{n_1-n_2} \, \frac{1}{2}  \, (\delta_{n_1}^0 + \delta_{n_2}^0) 
  \label{e:Keqc2Koc} \, .
\end{align}
Therefore, the time dependent power \eqref{e:PnK} at cell $n$ becomes\cite{min18}
\begin{equation}
P_{z,n}(t) \approx  \frac{c^2}{d} \sum_{s\in\ZZ} \sum_{n_2} \frac{1}{2} \, (\cV^s_{n} \, \cI^s_{n_2} - \cV^s_{n_2} \, \cI^s_{n} ) \, \sK^s_{n-n_2} \, . \label{e:1DMonoPower}
\end{equation}
We expect the accuracy of this approximation to deteriorate in nonlinear regime, 
because of the harmonic generation breaking the monochromatic regime.
Note that, from the group velocity $v_{\rmg} (\beta)=  \sum_{n\in \ZZ} \rmi \, n \, d \, \Omega^s_{n} \, \rme^{\rmi n \beta d}$, we have directly $\sK^s_{n}  =  d \sum_{n' \in \ZZ} (n' - n) \, \Omega^s_{n-n'}  \Omega^s_{n'} / c^2 $. 

Figure~\ref{f:Kn} displays the values of Eq.~\eqref{e:ksntransform} for the helix geometry.
The approximation \eqref{e:1DMonoPower} allows to write the power per cell 
using for Eq.~\eqref{e:sK} a substitute which depends only on the dispersion relation of the waveguide.

\begin{figure}
\centering
  \includegraphics[width=\columnwidth]{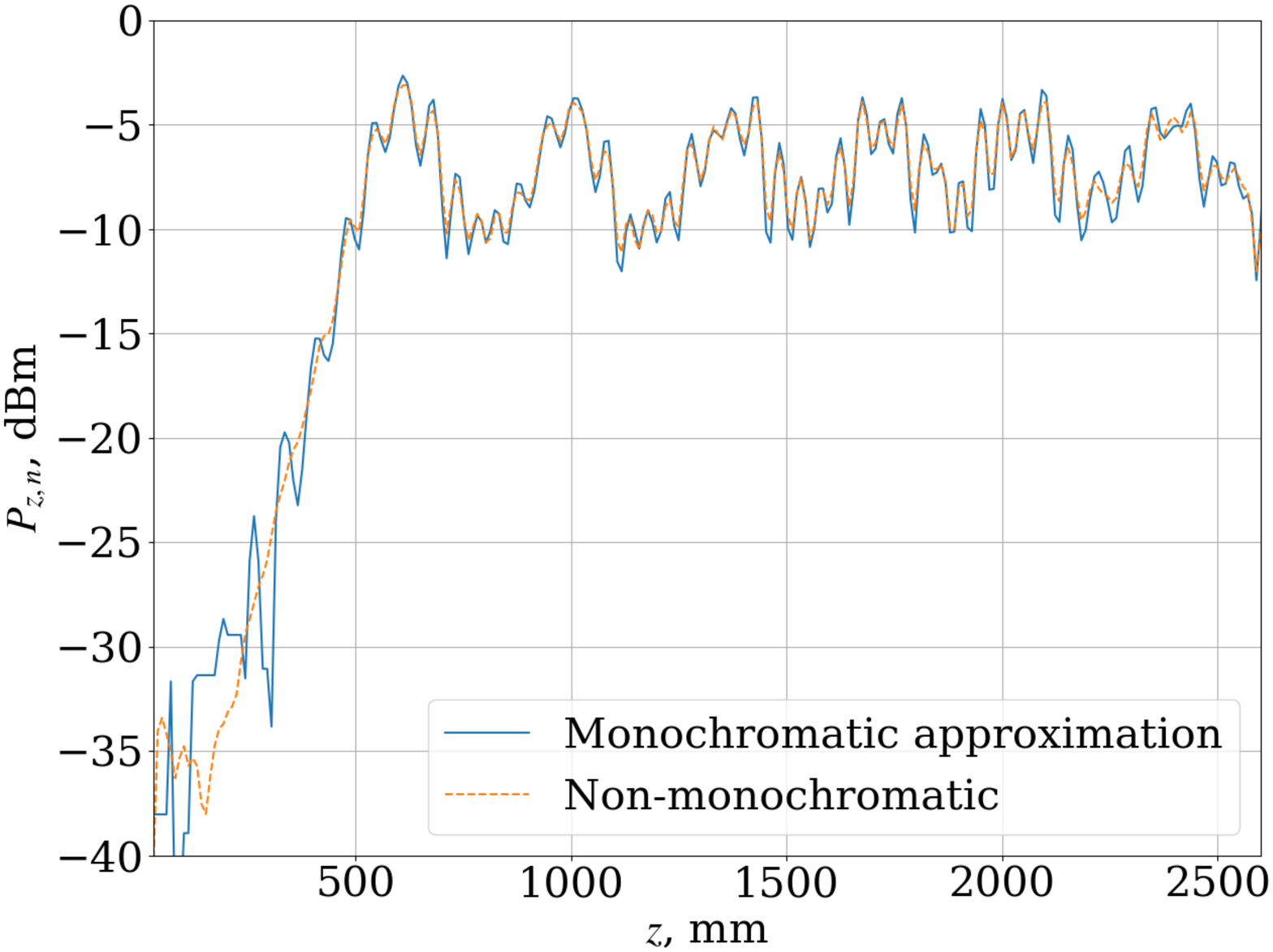}
  \caption{Electromagnetic power per cell $P_{z,n}(t)$ at time $t = 0.7$ \SI{}{\micro\second}. 
    The time amplitudes $\cV^s_n$, $\cI^s_n$ are obtained from the same simulation 
    as for Figs \ref{f:dimonte}, \ref{f:dimonteelec} and \ref{f:EMfields} for the 3 m long TWT.   
    The monochromatic expression is given by Eq.~\eqref{e:1DMonoPower}, 
    with Eq.~\eqref{e:ksntransform}. 
    The non-monochromatic expression is given by Eq.~\eqref{e:PnK}, 
    with Eqs~\eqref{e:finalKn1n2chroma} and \eqref{e:finalfrk}.
}
\label{f:powermono}
\end{figure}

\subsection{Explicit multi-frequency expression of the power}

The monochromatic approximation \eqref{e:1DMonoPower} of the power 
is efficient in linear regime for a wave with one frequency, as demonstrated in Ref.~\onlinecite{min18}.
However, when we reach the nonlinear regime, 
generation of strong harmonics occurs from the wave-particle interaction. 
In addition, the signal propagating in the waveguide can have several frequencies (multi-carrier), 
generating intermodulation products. 
Finally, oscillations and backward reflections can appear from the tube defects.
Therefore, the monochromatic approximation is a significant limitation for many applications. 

Now, we propose an explicit expression of the electromagnetic power for any multi-frequency field. 
To compute Eq.~\eqref{e:sK}, we need the $z$-dependence of the basis fields $\bcE^{s}_{n_1}, \bcH^{s}_{n_2}$, 
or equivalently of eigenfields $\bcE^{s}_{\beta_1}, \bcH^{s}_{\beta_2}$. 
In the sheath helix model ($s=0$), the spatial components are given by  Eqs~\eqref{e:erin}-\eqref{e:hzout}.
Therefore, Eq.~\eqref{e:kb1b2} becomes
\begin{align}
  \frk_{\beta_1, \beta_2} 
  = & \int_0^{2 \pi} \int^{a}_{0} \left( \cE^{\rmi *}_{r,\beta_1}  \cH^{\rm i}_{\theta,\beta_2} 
                                                     - \cE^{\rm i*}_{\theta,\beta_1}  \cH^{\rm i}_{r,\beta_2}  
                                                     \right) r \upd r \upd \theta 
   \nonumber \\
   &\quad \,  
      + \int_0^{2 \pi} \int^{\infty}_{a} \left( \cE^{\rm o*}_{r,\beta_1}  \cH^{\rm o}_{\theta,\beta_2} 
                                                            - \cE^{\rm o*}_{\theta,\beta_1}  \cH^{\rm o}_{r,\beta_2} 
                                                            \right) r \upd r  \upd \theta  
   \, .
\end{align}
Identities for the Bessel functions then yield
\begin{align}
  \frk_{\beta_1, \beta_2} 
  = & 
    \frac{2\pi a}{\Gamma^2_1 - \Gamma^2_2}  
    \Big(  \frac{\mu_0 \beta_2 \Omega_{\beta_1}}{\Gamma_1 \Gamma_2} 
             D^{\rm i *}_{\beta_1} D^{\rm i}_{\beta_2} 
           + \frac{\epsilon_0 \beta_1 \Omega_{\beta_2}}{\Gamma_1 \Gamma_2} 
              C^{\rm i *}_{\beta_1} C^{\rm i}_{\beta_2}  \Big) 
    \nonumber \\ 
   & \quad  \cdot 
     \Big( \Gamma_1 I_1(\Gamma_2 a) I_2(\Gamma_1 a)  
            -  \Gamma_2 I_1(\Gamma_1 a) I_2(\Gamma_2 a)  \Big) 
     \nonumber \\  &
   + \frac{2\pi a}{\Gamma^2_2 - \Gamma^2_1} 
      \Big(  \frac{\mu_0 \beta_2 \Omega_{\beta_1}}{\Gamma_1 \Gamma_2 } 
                D^{\rm o *}_{\beta_1} D^{\rm o}_{\beta_2}   
             + \frac{\epsilon_0 \beta_1 \Omega_{\beta_2}}{\Gamma_1 \Gamma_2} 
                C^{\rm o *}_{\beta_1} C^{\rm o}_{\beta_2} \Big) 
   \nonumber \\ 
   & \quad \cdot
   \Big( \Gamma_2 K_0(\Gamma_2 a) K_1(\Gamma_1 a) 
           -  \Gamma_1 K_0(\Gamma_1 a) K_1(\Gamma_2 a)  \Big) 
\label{e:finalfrk}
\end{align}
if $\beta_1 \neq \beta_2$, and its limit for $\beta_1 = \beta_2$ 
\begin{align}
  \frk_{\beta_1, \beta_1}
   = &  
    \frac{\pi a  \beta_1 \Omega_{\beta_1}}{\Gamma_1^2}  
      \Big(  \mu_0 | D^{\rm i}_{\beta_1} |^2 + \epsilon_0 | C^{\rm i}_{\beta_1} |^2 \Big)  
     \nonumber \\   & \quad
     \cdot \Big(I^2_1(\Gamma_1 a) - I_0(\Gamma_1 a) I_2(\Gamma_1 a)  \Big) 
   \nonumber \\   &
  + \frac{\pi a \beta_1 \Omega_{\beta_1}}{\Gamma_1^2} 
     \Big( \mu_0 | D^{\rm o}_{\beta_1} |^2 + \epsilon_0 | C^{\rm o}_{\beta_1} |^2 \Big) 
     \nonumber \\   & \quad
     \cdot \Big( K_0(\Gamma_1 a) K_2(\Gamma_1 a) - K_1^2(\Gamma_1 a) \Big)  
     \, ,
     \label{e:frkdiag}
\end{align}
with $\Gamma_1 = \sqrt{\beta_1^2 - \Omega_{\beta_1}^2/c^2}$, 
$\Gamma_2 = \sqrt{\beta_2^2 - \Omega_{\beta_2}^2/c^2}$ 
and the constants given by Eqs~\eqref{e:bondaBessel1}-\eqref{e:bondaBessel5}.
We can now obtain Eq.~\eqref{e:kn1n2kb1b2}, with a double Fourier transform on $\beta_1$ and $\beta_2$.
Therefore, we can express exactly the electromagnetic  power per cell within the sheath helix model.

Figure~\ref{f:Kbb} displays the values of $\frk_{\beta_1, \beta_2}$ 
as function of both wavenumbers for the helix geometry. 
Note that $\frk_{\beta_1, \beta_2}$ is real-valued. 
Equation \eqref{e:finalfrk} shows the two-frequency terms that were ignored in our monochromatic approximation.
Figure~\ref{f:Knn} displays the values of $\cK^{s}_{n_1, n_2}$ within the sheath helix model. 
We can see the discrepancies between the monochromatic approximation and the multi-frequency formulation.

Hence, we are now able to express the longitudinal electromagnetic power, 
Eq.~\eqref{e:PnK} per cell, or Eq.~\eqref{e:PVIK} as function of space, 
in the time domain without the monochromatic approximation.
Figure \ref{f:powermono} displays both the instantaneous monochromatic \eqref{e:1DMonoPower} 
and non-monochromatic \eqref{e:PnK} powers from the sheath helix model at a given time ;
for the same reasons as in Fig.~\ref{f:EMfields}, we combine the analytic kernels of the sheath model 
with the coefficients $\cV_n(t)$, $\cI_n(t)$ computed using the more realistic dispersion relation and impedance data. 
Both curves almost coincide since this simulation is performed for a main fundamental frequency 
while the harmonic generation remains too small to be significant.
Comparing to the time average power displays in Fig.~\ref{f:dimonte}, 
we see that both monochromatic and non-monochromatic powers of Fig.~\ref{f:powermono} are less smooth. 
Those large oscillations come from the bunching of electrons at a given time (see Fig.~\ref{f:dimonteelec}).

\section{Conclusion and perspectives}
\label{s:conclu}

In this paper, we present our reduction model, called the discrete model, which is an exact field decomposition valid for periodic structures, including TWTs. It allows to express the electromagnetic fields (see Eqs~\eqref{e:VsnEsn}-\eqref{e:IsnHsn}) in terms of time amplitudes $\cV_{n}, \cI_{n}$ and smooth shape functions $\bcE_{n}, \bcH_{n}$ depending on the structure geometry.

To compute the time amplitudes $\cV_{n}, \cI_{n}$, we combine the discrete model with a 1D multi-particle self-consistent Hamiltonian theory. From this, we build the \textsc{dimoha} algorithm to study the nonlinear wave-particle interaction in TWTs. Our algorithm is fast thanks to several characteristics including the DOF reduction from the discrete model and the space charge model. Our main result is the validation of our model against a full 3 m long experimental TWT. The simulations are performed in large nonlinear regime. Improvement of the particle dynamics is possible, especially for the space charge model; however, this may increase the computation time. The key advantages of \textsc{dimoha} are its swiftness compared
to alternative time domain approaches such as PIC codes and its explicit symplectic formulation which enforces the preservation of important physical features.

The shape functions $\bcE_{n}, \bcH_{n}$ are computed using a 3D helix model. 
From it, we can express the complete 3D propagation of the electromagnetic field inside the device.
The development of the helix version of the discrete model, presented in this paper, 
can be used as a canvas for investigating another 3D geometry with the discrete model.
Such models pave the way to 3D interaction codes.
We are able to reconstruct the 3D field from the time amplitudes $\cV^{s}_{n}, \cI^{s}_{n}$ 
computed with our 1D simulations.

The helix version of the discrete model is also used to express the electromagnetic power.
Generally, this is the average power for one frequency (monochromatic harmonic power) that is computed.
We propose an explicit formulation of the electromagnetic power in the time domain and in non-monochromatic regime. 
We use the sheath helix model, but it is possible to change the geometry following the same development as ours.

This work opens several perspectives.
Now, we have a benchmarked tool with reasonable computational times that can be used to investigate nonlinear effects, like the coupling of multiple waves or chaos.
The current version of \textsc{dimoha} can be improved,  
in particular the space charge model. 
A version of \textsc{dimoha} for the design of industrial TWTs is under development.
We are also working on a comparison\cite{ali21} between time-domain pseudospectral models\cite{con04} and \textsc{dimoha}.
Finally a 2D or 3D version of \textsc{dimoha}, with the helix geometry, is under consideration.

\acknowledgments

The authors thank M. C. de Sousa and anonymous reviewers for their critical reading of the manuscript,
and M. Tacu, D. Tordeux and members of \'Equipe Turbulence Plasma for fruitful discussions.
Centre de Calcul Intensif d'Aix-Marseille is acknowledged for granting access to its high performance computing resources. 
The authors express their condolences to the family and friends of Professor Sergey Petrovich Kuznetsov, the instigator of the discrete model, who passed away in 2020 at the age of 69.

\section*{Data availability}

The data that support the findings of this study are available from the corresponding author upon reasonable request.

\section*{References}

%
\end{document}